\documentclass[aps,pre,twocolumn,epsfig,floats]{revtex4}
\usepackage{amsmath}
\usepackage{color}
\usepackage{graphicx}
\usepackage{epsfig}
\usepackage{amsfonts}
\usepackage{mathrsfs}
\usepackage{mathtools}
\usepackage{amsmath}
\usepackage{float}
\usepackage{physics}
\usepackage{hyperref}
\hypersetup{
  colorlinks=true,
  citecolor=blue,
  linkcolor=blue,
  urlcolor=blue}

\begin{document}

\title{Quantum transitions, ergodicity, and quantum scars in the coupled top model}

\author{Debabrata Mondal, Sudip Sinha, and S. Sinha}
\affiliation{Indian Institute of Science Education and
	Research-Kolkata, Mohanpur, Nadia-741246, India}

\begin{abstract}
We consider an interacting collective spin model known as coupled top (CT), exhibiting a rich variety of phenomena related to quantum transitions, ergodicity, and formation of quantum scars, discussed in [Phys. Rev. E {\bf 102}, 020101(R) (2020)]. In this work, we present a detailed analysis of the different type of transitions in CT model, and find their connection with the underlying collective spin dynamics. Apart from the quantum scarring phenomena, we also identify another source of deviation from ergodicity due to the presence of non-ergodic multifractal states. The degree of ergodicity of the eigenstates across the energy band is quantified from the relative entanglement entropy as well as multifractal dimensions, which can be probed from non-equilibrium dynamics. Finally, we discuss the detection of non-ergodic behavior and different types of quantum scars using `out-of-time-order correlators', which has relevance in the recent experiments.
\end{abstract}

\date{\today}
\maketitle
\section{Introduction}
The study of the ground state, as well as excited states of interacting quantum systems, has become important for understanding various phenomena such as quantum phase transition (QPT), entanglement properties and more importantly, their ergodic behavior in non-equilibrium dynamics \cite{Subir_QPT,Amico_RMP,Kafri,Polkovnikov_colloquium,Gogolin}. The QPT across a critical point leads to the ordering of the ground state, which is also reflected from the critical behavior of physical quantities at zero temperature \cite{Subir_QPT}. In recent years, another type of quantum transition has been observed in collective models, where the excited states change their characteristics at a critical energy density corresponding to the singularities of density of states. Such transition is termed as excited state quantum phase transition (ESQPT) \cite{ESQPT_topical_review,Cejnar1,Caprio,Cejnar2,dicke_ESQPT,dicke_ESQPT2,L_Santos}, however its connection with thermodynamic behavior of physical quantities deserves further attention \cite{ESQPT_topical_review,dicke_ESQPT2}. The non-equilibrium dynamics of a closed quantum system reveals various interesting phenomena related to ergodicity of the system, where the excited states also play a crucial role \cite{Polkovnikov_colloquium,Moore_review}. In recent years, the ultracold atomic system has become a test bed to study the non-equilibrium properties of quantum many body systems \cite{Altman_lectures,ultracold_atoms_dynamics}. Certain many body systems after a quench evolve to a steady state corresponding to a generalized thermal distribution \cite{Markos_Rigol,Marcos_Rigol_GGE,Langen,Gring,Caux,Dutta}. To understand thermalization of closed quantum systems, the eigenstate thermalization hypothesis (ETH) has been proposed, which attempts to explain ergodicity at the level of individual eigenstates \cite{Deutsch,Srednicki}. Moreover, its connection with random matrix theory (RMT) has been extensively explored in different quantum systems \cite{Izrailev_1,Izrailev_2,Kafri}. 

On the contrary, there are examples where thermalization is absent, such as systems exhibiting many body localization (MBL) \cite{MBL_1,MBL_2,MBL_3}. Apart from MBL, the presence of non-ergodic multifractal states \cite{Altshuler,Deng,Fazio_bhm} can also lead to the deviation from ergodicity and anomalous thermalization \cite{Luitz}. Tuning the parameters of quantum many body systems can change the  nature of eigenstates, which in turn leads to the transition from MBL to ergodic or non-ergodic phases \cite{I_Bloch,Schreiber,Polkovnikov,Abanin_1,Arnab_Das,Altshuler_2,S_Ray_1}. Recently, such non-ergodic and ergodic evolution has also been studied in quantum circuits \cite{quantum_circuit,quantum_cellular_automaton}.

Beside the above mentioned cases, there are other sources of deviation from ergodicity, such as many body quantum scarring phenomena, which has recently been observed in an experiment on a chain of ultracold Rydberg atoms \cite{Bernien}. It has been found that a special choice of initial state exhibits periodic revival phenomenon and fails to thermalize, which has been attributed to many body quantum scar (MBQS) \cite{Turner,Motrunich,emergent_SU2,M_Lukin1,abanin,Silva_scar}. Such phenomena has also been analyzed theoretically in other interacting quantum models \cite{spin1,onsager_scars,correlated_hopping,fracton,
optical_lattice,K_saito,M_Lukin2,sengupta_scars,AKLT,AKLT2,sinha1,mondal}. Originally, quantum scars in non interacting quantum system were identified as reminiscence of unstable classical orbits in a chaotic stadium \cite{Heller}. However, such connection of unstable dynamics with MBQS in an interacting quantum system is not obvious and deserves further attention \cite{abanin,M_Lukin1,sinha1,mondal}.

Unlike the classical systems, the route to ergodicity and its deviation in closed quantum systems remains a challenging issue. In a classical system, the ergodicity due to phase space mixing arises from irregular behavior of chaotic trajectories, 
whereas such picture is unclear in a quantum system due to the absence of phase space trajectories. However, the connection between ergodicity and underlying chaotic dynamics of its classical counterpart has been explored in certain systems \cite{Altland_Haake,Kicked_dickie,KCT_mondal}. Moreover, the underlying dynamical instability in a quantum system can be detected by a newly developed tool known as ‘out-of-time-order correlator’ (OTOC), which allows us to investigate the route to ergodicity and its connection with underlying chaos \cite{stanford2,maldacena,K_hashimoto,Rozenbaum,Garttner1,butterfly_effect,Swingle1,
NMR,Trapped_ion,Fazio_otoc,A_M_rey2,Santos_otoc,
OTOC_instability_LMG,Garcia_mata,lakshminarayan,sray1,sudip_otoc}. 
In this context, it is also important to explore the connection between formation of MBQS in an interacting many body system and the underlying unstable dynamics, which has not been properly established. Such correspondence with classical description can elucidate the ergodic behavior of closed quantum systems and its deviation due to the formation of quantum scars, which is one of the main objectives of the present work.

In this work, we consider an interacting collective spin model known as coupled top (CT) model, exhibiting a variety of rich phenomena, which includes different type of quantum transitions, ergodic behavior and formation of quantum scars. We discuss the above mentioned phenomena in details and explore their connection with the underlying classical dynamics. The CT model undergoes a QPT at a critical coupling, above which the onset of chaos occurs in an intermediate range of coupling strength, where its ergodic behavior is investigated in details. Even in the chaotic regime, we identify the sources of deviation from ergodicity due to the presence of non-ergodic multifractal states and quantum scars, which can be detected from their dynamical signatures. Moreover, we elucidate the mechanism behind the formation of quantum scars arising from the unstable fixed points and periodic orbits. This work is an extension of our previous work \cite{mondal}, where we present an elaborate discussion and results related to quantum transitions, ergodic properties and quantum scarring phenomena in the CT model. 

The paper is organized as follows. In Sec.\ref{Classical}, we describe the coupled top (CT) model and analyze it classically to obtain the steady states and their stability. Next, we study the various type of quantum transitions exhibited by this model and discuss their connection with the classical steady states in Sec.\ref{quantum_transition_CT}. It undergoes a quantum phase transition (QPT) as well dynamical transition at a critical coupling, which is discussed in subsection \ref{QPT_DT}, and the occurrence of excited state quantum phase transition (ESQPT) above the critical coupling is discussed in subsection \ref{ESQPT_CT}. In Sec.\ref{Onset of chaos}, we investigate the onset of chaos both classically as well quantum mechanically, which are presented in subsections \ref{Lyapunov} and \ref{Spectral_Statistics} respectively. 
Next, we investigate the ergodic properties of the eigenstates in Sec.\ref{Ergodic} and discuss their multifractal behavior in subsection \ref{multifractality}. The quantum dynamics of CT model is presented in Sec.\ref{quantum_dynamics}, and its classical quantum correspondence is explored in subsection \ref{classical_quantum_correspondence}. The manifestation of degree of ergodicity in non-equilibrium dynamics and its detection by using the `out-of-time-order correlators' (OTOC), are discussed in subsections \ref{non_ergodic_dynamics} and \ref{OTOC dynamics} respectively. The quantum scarring phenomena in CT model is investigated in Sec.\ref{Quantum scar}, where we identify the two types of scars arising from unstable steady states and unstable periodic orbits, which are presented in subsection \ref{Scar_FP} and \ref{Scar_periodic_orbit}. In subsection \ref{Dynamical_signeture_scar}, we discuss the detection of quantum scars from their dynamical signature using OTOC technique. Finally, we summarize the results and conclude in Sec.\ref{Conclusion}. In appendix \ref{Holstein_primakoff}, we outline the Holstein Primakoff approximation to obtain the excitation energies and spin fluctuations at QPT. In appendix \ref{effective_potential}, we derive the effective potential corresponding to CT model, describing the QPT and ESQPT. The signature of QPT from multifractal dimensions is presented in appendix \ref{multifractal_dimension_QPT}. The integrability of CT model at an extreme coupling strength is discussed in appendix \ref{integrability_large_coupling}.
\section{Model and classical analysis} 
\label{Classical}
%%%%%%%%%%%%%%%%% phase portrait  and classical analysis Fig:1  %%%%%%%%%%%%%%%%%%%%
\begin{figure*}
	\centering
	\includegraphics[height=7cm,width=18cm]{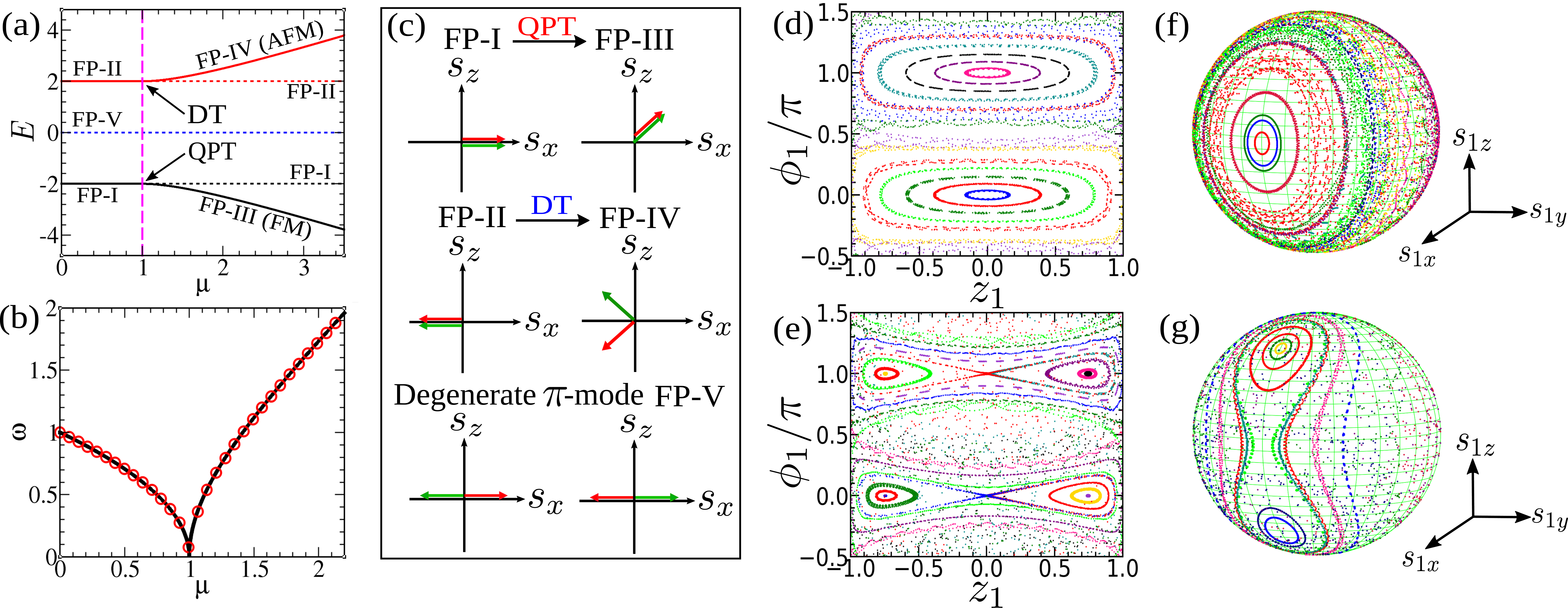}
	\caption{ (a) Various steady states with different energies as a function of $\mu$. Solid (dotted) lines represent stable (unstable) branches and vertical pink dashed line indicates the critical coupling $\mu_c$. (b) Small amplitude oscillation frequency $\omega$ as a function $\mu$ of the steady states corresponding to ground state (black solid line) and highest excited state (red circles), where vanishing of $\omega$ at $\mu_c=1$ signifies QPT and dynamical transition. (c) Schematic diagram of spin configuration for different steady states. Phase portraits ((d) and (e)) and trajectories on Bloch sphere ((f) and (g)) exhibiting QPT and dynamical transition respectively. For (d),(f) $\mu=0.5< \mu_c$ and (e),(g) $\mu=1.5>\mu_c$.}
	\label{phase-portrait}
\end{figure*}
%%%%%%%%%%%%%%%%%%%%%%%%%%%%%%%%%%%%%%%%%%%%%%%%%%%%%%%%%%%%%%%%%%%%%
The coupled top (CT) model \cite{chaos,kicked_spin1,classical_analysis,ct_EE,S_Ray_D_sen,mondal} describes the dynamics of two large spins of equal magnitude $S$ interacting ferromagnetically with each other, analogous to transverse field Ising model \cite{Ising_1}. This collective model is described by the following Hamiltonian,
	\begin{eqnarray}
	\hat{\mathcal{H}}&=&-\hbar\, \omega_0\,(\hat{S}_{1x}+\hat{S}_{2x})-\,\frac{\mu}{S}\,\hat{S}_{1z}\,\hat{S}_{2z}
	\label{Coupled Top}
	\end{eqnarray}
where $\hat{S}_{ia}$ ($a = x,y,z$) represent the spin operators corresponding to two spins $(i = 1, 2)$. The first term in Eq.\eqref{Coupled Top} describes the precession of two non interacting spins around the $x$-axis with angular frequency $\omega_0$, while the second term denotes ferromagnetic interaction between them with coupling strength $\mu$. In rest of the paper, we scale energy (time) by $\omega_0 (1/\omega_0 )$ and set $\hbar = 1$, unless otherwise mentioned.

The collective nature of the spins, in the limit $S\rightarrow \infty$ allows us to study the model classically. In this limit, the quantity $\hat{s}_{ia}=\hat{S}_{ia}/S$ can be written in terms of the classical spin vector,
\begin{eqnarray}
\vec{s_i}\equiv(s_{ix},s_{iy},s_{iz})=(\sin\theta_i\cos\phi_i,\sin\theta_i\sin\phi_i,\cos\theta_i)
\label{Classical spin vector}
\end{eqnarray}
 where $\phi_i$ and $\theta_i$ represent the orientation of the spins. Alternatively, the dynamics can be  studied in phase space using the canonically conjugate variables $\{\phi_i,z_i=\cos\theta_i\}$ and in terms of these collective variables $\{z_1,\phi_1,z_2,\phi_2\}$, the classical Hamiltonian can be rewritten as, 
\begin{eqnarray}
\mathcal{H}_{cl}=-\sqrt{1-{z_1}^2}\cos\phi_1-\sqrt{1-{z_2}^2}\cos\phi_2-\mu \,z_1 z_2
\label{classical_Hamiltonian_coupled_top}
\end{eqnarray}
where the Hamiltonian $\mathcal{H}_{cl}$ and the classical energy $E$ are scaled by spin magnitude $S$, therefore becoming an intensive quantity, independent of $S$. In general, we denote the collective variables as $\mathbf{X}=\{z_1,\phi_1,z_2,\phi_2\}$, which represents the phase space point. The classical dynamics is described by the following equations of motion (EOM),
\begin{subequations}
\begin{eqnarray}
\dot{\phi_i}&=&\frac{z_i\cos\phi_i}{\sqrt{1-{z_i}^2}}-\mu z_{\bar{i}} \label{classical_EOM-phi} \\
\dot{z_i}&=&-\sqrt{1-{z_i}^2}\,\,\sin\phi_i
\label{classical_EOM-z}
\end{eqnarray}
\label{EOM}
\end{subequations}
where $\bar{i} \neq i$. 
To capture the overall dynamical behavior, we first analyze the fixed points (FP) and their stability, varying the coupling strength $\mu$. The FPs describe the steady states, which can be obtained by equating the LHS of Eq.(\ref{EOM}) to zero and are denoted by $\{z_1^*,\phi_1^*,z_2^*,\phi_2^*\}$.  To understand the stability of the steady states, we perform the linear stability analysis in the presence of small fluctuations around the FPs, which evolves as,
$\delta\mathbf{X}(t)=\delta\mathbf{X}(0) e^{i\omega t}$. The frequency $\omega$ of the small amplitude oscillation, is given by,
\begin{eqnarray}
{\omega}_{\pm}^2=\frac{1}{2}\left(\text{A}_1+\text{A}_2\pm \sqrt{\left(\text{A}_1-\text{A}_2\right)^2+4\mu ^2\frac{\cos\phi_1^* \cos\phi_2^*}{\sqrt{\text{A}_1\text{A}_2}}}\right)
\end{eqnarray}
where $\omega_{\pm}$ are the frequencies of two collective modes and $\text{A}_i=1/(1-{z_i}^{*2})$ for $i = 1,2$. The only possible stable FPs in these Hamiltonian (conservative) systems are `centers' and their stability is ensured if the oscillation frequencies are real, equivalently $\lambda_{\text{I}}=\Im(\omega)=0$. On the other hand, the unstable FPs correspond to `saddles', which arises when $\lambda_{\text{I}}\neq0$ \cite{Strogartz}.

The steady states and their nature with increasing coupling strength are summarized in Fig.\ref{phase-portrait}(a), where we observe a characteristic change at a critical coupling strength $\mu_c=1$. The nature of the FPs can be characterized by the Ising symmetry of the Hamiltonian, which remain invariant under $s_{iz}\rightarrow -s_{iz}$.
For  $\mu < \mu_c$, we find the following symmetry unbroken stable FPs, which are represented by,
\begin{eqnarray}
&&(\text{I}):\,\, z_1^*=0,\phi_1^*=0;z_2^*=0,\phi_2^*=0\\
&&(\text{II}):\,\, z_1^*=0,\phi_1^*=\pi;z_2^*=0,\phi_2^*=\pi
\end{eqnarray}
For FP-I (FP-II), both spins are aligned to positive (negative) $x$-axis, and have zero magnetization along $z$-axis, with energy $E=\mp 2$. Both FP-I and II become unstable at $\mu=\mu_c$ and undergo pitchfork bifurcation, giving rise to two pairs of stable symmetry broken steady states, which are given by,
\begin{eqnarray}
&&(\text{III}):\,\, z_1^*=z_2^*=\pm\sqrt{1-1/\mu^2},\phi_1^*= \phi_2^*=0\\
&&(\text{IV}):\,\, z_1^*=-z_2^*=\pm\sqrt{1-1/\mu^2},\phi_1^*=\phi_2^*=\pi
\end{eqnarray}
where (FP-IV) FP-III represents (anti)ferromagnetic state with energy $E=\pm(\mu+1/\mu)$, which corresponds to the maximum (minimum) energy of the classical Hamiltonian in Eq.\eqref{classical_Hamiltonian_coupled_top}. Both above and below the critical coupling  $\mu_c$, the stable FPs corresponding to maximum and minimum energy states yield same  frequencies of small amplitude oscillations, although their spin configurations are different. The oscillation frequencies for symmetry unbroken (FP-I, FP-II) and broken (FP-III, FP-IV) states are given by,
\begin{eqnarray}
\omega_{\pm} &=&
\begin{dcases*} 
\sqrt{1\pm \mu} & for  $\mu < \mu_c$ \\ 
\sqrt{\mu^2\pm 1} & for $\mu \ge \mu_c$
\end{dcases*} \label{Excitation frequency}
\end{eqnarray} 
It is interesting to note that, the oscillation frequency $\omega_-$ of the lower mode vanishes at $\mu_c$ for both the bifurcations, as shown in Fig.\ref{phase-portrait}(b)).

In addition, there exists another degenerate pair of unstable FPs represented by (V) $\{z_1^*=z_2^*=0,\phi_1^*=\pi,\phi_2^*=0\}$ and $\{z_1^*=z_2^*=0,\phi_1^*=0,\phi_2^*=\pi\}$ with energy $E=0$. We denote this as `$\pi$-mode', since the relative angle between the spins is $\pi$.
The instability exponent $\lambda_{\text{I}}$ of the `$\pi$-mode' (FP-V) is given by,
\begin{eqnarray}
\lambda_{\text{I}}&=&\sqrt{\frac{(\mu^2+1)^\frac{1}{2}-1}{2}}  \label{Excitation frequency-pi}
\end{eqnarray} 
which increases with the coupling strength $\mu$.

The schematic, explaining the spin configurations corresponding to the above FPs are given in Fig.\ref{phase-portrait}(c). To clearly visualize the fixed point structure and their bifurcation, we obtain the dynamics, by solving the EOM given in Eq.\eqref{EOM} and plot it over phase portrait in $z_1-\phi_1$ plane and Bloch sphere (see Fig.\ref{phase-portrait}(d-g)). It can be observed from Fig.\ref{phase-portrait}(d,f), for small coupling strength $\mu<\mu_c$, regular phase space trajectories are formed around the symmetry unbroken phases, whereas, above the critical point, the phase portrait contains regular trajectories around the symmetry broken FPs. However, the intermediate region of phase space is filled with irregular trajectories (see Fig.\ref{phase-portrait}(e)), indicating the onset of chaos, which is discussed in Sec.\ref{Onset of chaos}.
\section{Quantum transitions in coupled top model}
\label{quantum_transition_CT}
In this section, we discuss different types of quantum transitions exhibited by the CT model and identify their signature from the behavior of different physical quantities. The connection between these transitions with the fixed point structure and their bifurcation, obtained from classical analysis are also studied. 
\subsection{Quantum phase transition and Dynamical transition}
\label{QPT_DT}
\begin{figure}
		\centering
		\includegraphics[height=11cm,width=8.8cm]{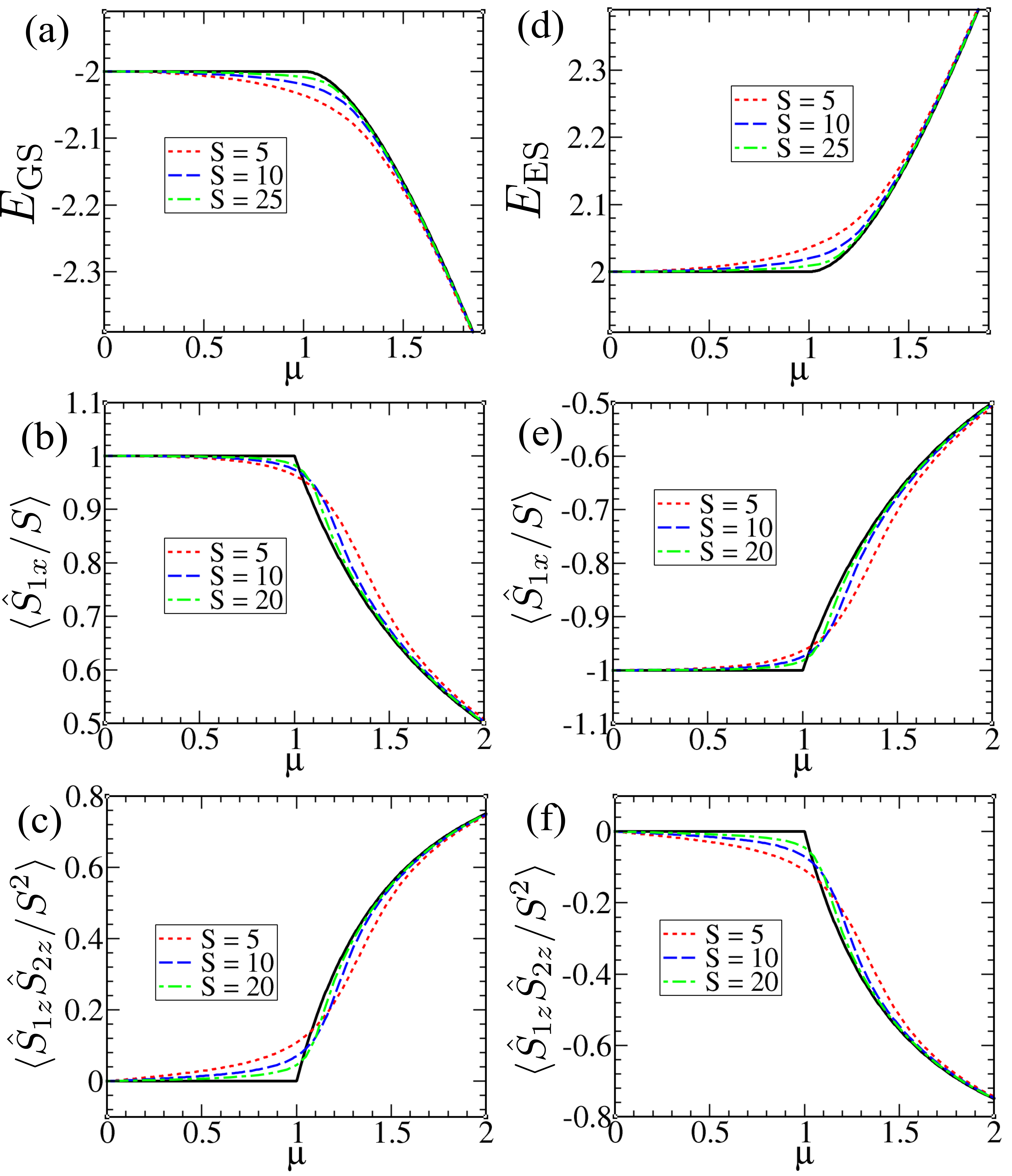}
		\caption{Signature of QPT and dynamical transition:  Variation of (a) ground state energy $E_{\rm GS}$, (b) $\langle \hat{S}_{1x}/S \rangle$, (c) $\langle \hat{S}_{1z}\hat{S}_{2z}/S^2 \rangle$ as a function of coupling strength $\mu$. The second column (d,e,f) represents variation of same physical quantities as in (a,b,c) with $\mu$, corresponding to highest excited state, exhibiting dynamical transition. The black solid line in all the figures denote the classically obtained analytical result of the respective quantities.}
		\label{fig:2}
\end{figure}

As seen from the classical analysis, the bifurcation of the steady state FP-I to FP-III breaks the Ising symmetry, which gives rise to a ferromagnetic ordering of spins in the ground state (GS). Such bifurcation of the minimum energy steady state is identified as quantum phase transition (QPT) leading to the ordering of GS.
Similarly, the bifurcation of FP-II to FP-IV corresponds to a dynamical transition  associated with the highest energy excited state, acquiring an antiferromagnetic ordering. Such type of dynamical transition has also been observed in Bose-Josephson junction \cite{shenoy,Oberthaler}.
Quantum mechanically, both the QPT and dynamical transition can be identified from the changes in the relevant physical quantities computed for the GS and highest excited state respectively, which are obtained by diagonalizing the Hamiltonian given in Eq.\eqref{Coupled Top}. It is interesting to note that, both QPT and dynamical transition are related, since the transformation $\hat{S}_{x}\rightarrow -\hat{S}_{x}$ and $\mu \rightarrow -\mu$ changes the ground state of ferromagnetic coupled top to the excited state of the same model with antiferromagnetic interaction, which explains the antiferromagnetic ordering of the excited state after dynamical transition. Both QPT and dynamical transition can be captured by the characteristic changes in different physical quantities obtained for GS and excited state respectively, across $\mu=\mu_c$ (see Fig.\ref{fig:2}). As a consequence of the symmetry between the GS and excited state, the complementary behavior of the physical quantities (as shown in Fig.\ref{fig:2}) can be observed for QPT and dynamical transition, particularly, from the expectation value of $ \hat{S}_{1z} \hat{S}_{2z}/S^2 $, which captures the ferromagnetic and antiferromagnetic ordering due to the respective transitions (see Fig.\ref{fig:2}(c),(f)).
%%%%%%%%%%%%%%%%%%%%%%%% FIG:2 QPT and DT %%%%%%%%%%%%%%%%%%%%%%%%%%%%%%%%%%%%%
%%%%%%%%%%%%%%%%%%%%%%%%%%%%%%%%%%%%%%%%%%%%%%%%%%%%%%%%%%%%%%%%%%%

In the $S\rightarrow \infty$ limit, the QPT can be analyzed within Holstein Primakoff (HP) approximation \cite{HPT}, which has been discussed in details in appendix \ref{Holstein_primakoff}. The lowest excitation energy obtained from the HP method is in agreement with the small amplitude oscillation frequency (as shown in Fig.\ref{phase-portrait}(b)), which vanishes as $\sim \sqrt{|\mu-\mu_c|}$, signifying the mean field like behavior of QPT. As a consequence, the spin fluctuation diverges at the critical point, which can also be captured from the HP approximation. Moreover, within such semiclassical description, the spin system can equivalently be described in terms of the effective potential, which is discussed in appendix \ref{effective_potential}. For $\mu<\mu_c$, the effective potential has a single minima corresponding to the symmetry unbroken phase, which changes its shape to a double well structure, clearly exhibiting the symmetry breaking across the critical point.

Unlike the HP approximation (with $S\rightarrow \infty$), for full quantum mechanical computation with finite size Hilbert space, the true phase transition with symmetry broken phase is absent. However, after the transition, the energy gap between the consecutive even-odd parity eigenstates (symmetry preserved) is suppressed exponentially with system size. Although, the finite size effect masks the sharp changes of the physical quantities at the critical point, however with increasing $S$, they approach the classically obtained analytical results, exhibiting such sharp features, as depicted in Fig.\ref{fig:2}. For comparison of the numerical results with the classical ones in Fig.\ref{fig:2}, we have appropriately scaled the different physical quantities and quantum mechanical energies $\mathcal{E}_{\nu}$ by $S$, to make them intensive in nature. Note that, the classical energy is equivalent to quantum mechanical energy density ($E\equiv \mathcal{E}_{\nu}/S$).
\subsection{Excited state quantum phase transition}
\label{ESQPT_CT}
\begin{figure}
	\centering
	\includegraphics[height=7.4cm,width=8.8cm]{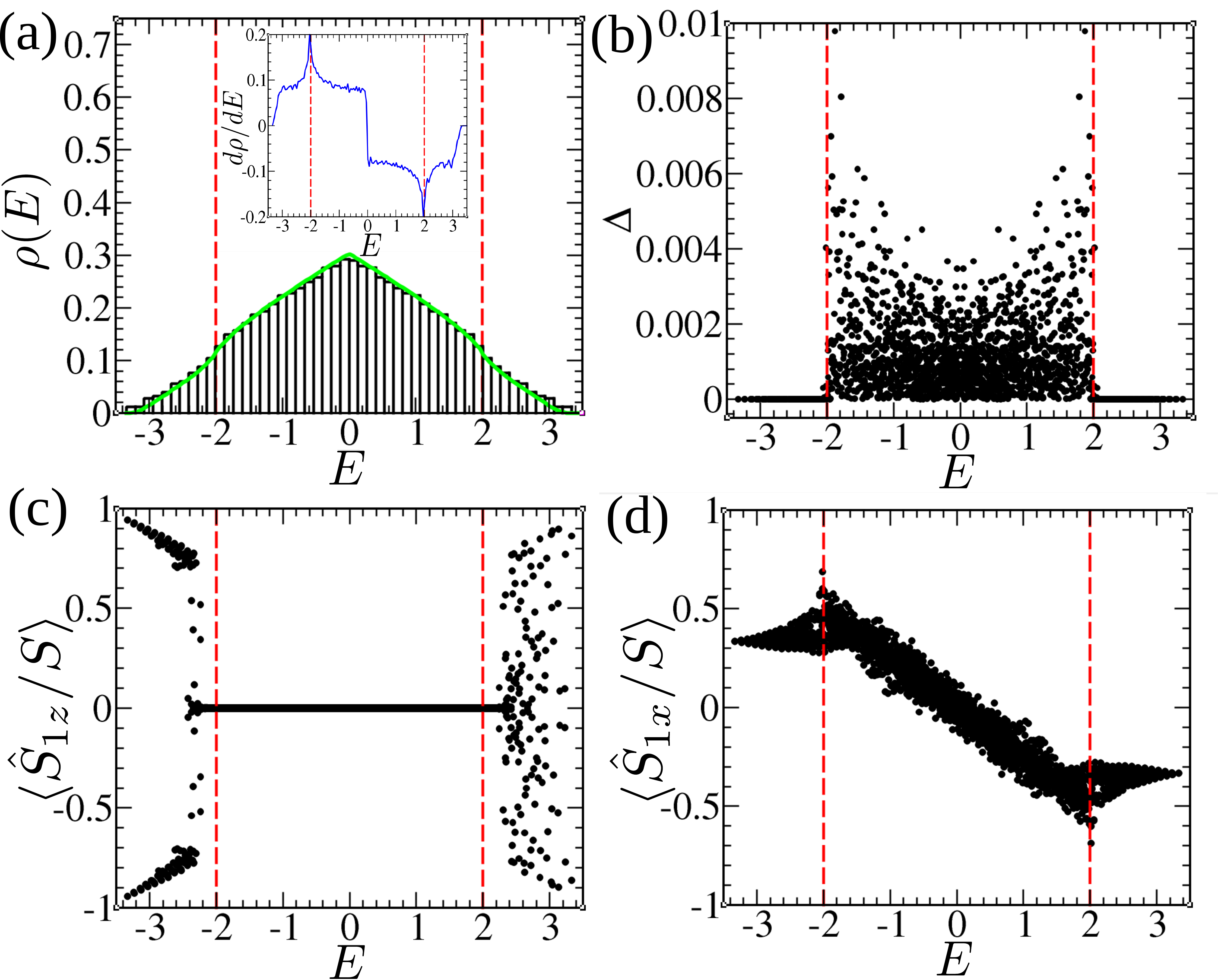}
	\caption{Manifestation of ESQPT:(a) Density of states $\rho(E)$ and its derivative $d\rho /dE$ shown in the inset, (b) pair gap $\Delta_{\nu}$, (c) $\langle\hat{S}_{1z}/S\rangle$ and (d) $\langle\hat{S}_{1x}/S\rangle$ as a function of energy density $E$ at $\mu=3$. The red dashed lines marked at $E=\pm2$ indicate the critical energy densities corresponding to ESQPT. For quantum calculations, we choose $S=30$ in this and the rest of the figures,  unless otherwise mentioned.}
	\label{ESQPT}
\end{figure}

Apart from QPT and dynamical transition, the CT model exhibits another kind of transition known as excited state quantum phase transition (ESQPT) \cite{ESQPT_topical_review,Cejnar1,Caprio,Cejnar2,dicke_ESQPT,dicke_ESQPT2,L_Santos}, which describes the characteristic change in the nature of excited eigenstates across certain critical energy density. Both QPT and dynamical transition are accompanied by ESQPT, which occurs above critical coupling strength $\mu_c$, at a critical energy density $E_c=\mp 2$, corresponding to unstable symmetry unbroken steady states FP-I and II respectively.
The essential feature of ESQPT is the singular behavior of density of states (DOS) at the critical energy density $E_c$. 
From the classical Hamiltonian $\mathcal{H}_{cl}$ (Eq.\eqref{classical_Hamiltonian_coupled_top}), DOS can be written as, 
\begin{small}
\begin{eqnarray}
%\textstyle
\rho(E) = \frac{\mathcal{C}}{4\pi^2}\int_{-1}^{1}\int_{-1}^{1}\int_{0}^{2\pi}\int_{0}^{2\pi}\delta(E-\mathcal{H}_{cl})\, d\phi_1 d\phi_2 dz_1 dz_2 
\label{DOS}
\end{eqnarray}
\end{small}
Where $\mathcal{C}=1/4$  keeps the normalization condition $\int \rho(E)\,dE = 1$, which is useful for the comparison of the semiclassical result with the exact quantum mechanical DOS, obtained for a system with finite $S$. The DOS obtained from quantum mechanical energy spectrum, is in good agreement with the semiclassical result, as shown in Fig.\ref{ESQPT}(a). For $\mu>\mu_c$, the derivative of the semiclassical DOS shown in the inset of Fig.\ref{ESQPT}(a) reveals the singularities at the critical energy densities indicating ESQPT. Physically, above $\mu_c$,  FP-I and II at critical energy densities $E_c=\mp2$
separate the symmetry unbroken states within the range $-2 < E < 2$ from the symmetry broken states in the energy range $E_{\text{GS}} < E < -2 $ and $ 2 < E < E_{\text{ES}}$ corresponding to QPT and dynamical transition respectively. The ESQPT corresponding to QPT can be elucidated in terms of the effective potential, as discussed in appendix \ref{effective_potential}. Above the critical point, the effective potential takes a double well structure, where the energy of the barrier height at $E=-2$ corresponds to the critical energy density $E_{c}$ of the ESQPT, which clearly separates the ferromagnetically ordered symmetry broken states ($E<E_c$) from the unbroken ones ($E>E_c$). To understand the above mentioned behavior of the eigenstates, we compute the energy gap between consecutive even and odd parity states, defined as $\Delta_{\nu} =\mathcal{E}_{2\nu}-\mathcal{E}_{2\nu-1}$, which vanishes exponentially with increasing $S$ for quasi degenerate states \cite{symm_breaking_1,symm_breaking_2} with $|E|>|E_c|$ revealing their symmetry broken nature (see Fig.\ref{ESQPT}(b)). Moreover, as shown in Fig.\ref{ESQPT}(c,d), the expectation values of the observables like $\langle \hat{S}_{1z}/S\rangle$ and $\langle \hat{S}_{1x}/S\rangle$ clearly separate the symmetry unbroken states from symmetry broken sector with a significant
change at $E_c = \mp2$ indicating ESQPT. The critical energy densities corresponding to ESQPT separating the symmetry broken and unbroken states, are also important for the ergodic nature of the eigenstates, which is discussed in the later part of this work.
%%%%%%%%%%%%%%%%%%%%%%%% FIG:3 ESQPT %%%%%%%%%%%%%%%%%%%%%%%%%%%%%%%%%%%%
%%%%%%%%%%%%%%%%%%%%%%%%%%%%%%%%%%%%%%%%%%%%%%%%%%%%%%%%%%%%%%%%%%%%%%%%%
\section{Onset of chaos }
\label{Onset of chaos}
In this section, we study the onset of chaos in the CT model, from the phase space dynamics and its reflection in the spectral statistics. In the weak coupling limit, the model exhibits an integrable structure, which represents two almost non interacting spins. The deviation from such regular dynamics occurs with increasing the coupling strength $\mu$, which is evident from the classical phase portrait shown in Fig.\ref{phase-portrait}(e,g). Above the critical coupling strength $\mu_c$, the regular region around the stable FPs shrinks and remaining part is filled up with chaotic trajectories, revealing a mixed phase space structure in the intermediate regime of coupling. As $\mu$ increases, the chaotic region in phase space continues to grow and it is eventually filled with irregular trajectories, at a fixed classical energy. This crossover from regular to chaotic regime can be captured from the Poincar\'{e} section corresponding to a fixed energy. However, it is important to note, in the extreme limit with $\mu\gg1$, the integrability of the scaled Hamiltonian $\hat{\mathcal{H}}/\mu$ can be recovered, which exhibits regular orbits around $z$-axis. Such integrable structure of the Hamiltonian  and regular dynamics for $\mu \gg 1$ is discussed in details in appendix \ref{integrability_large_coupling}.
\subsection{Poincar\'{e} section and Lyapunov exponent}
\label{Lyapunov}
%%%%%%%%%%%%%%%%% Fig:4 poincare sections and Lyapunov exponents %%%%%%%%%%%%%%%%%%%
\begin{figure}[H]
	\centering
	\includegraphics[height=7.4cm,width=8.7cm]{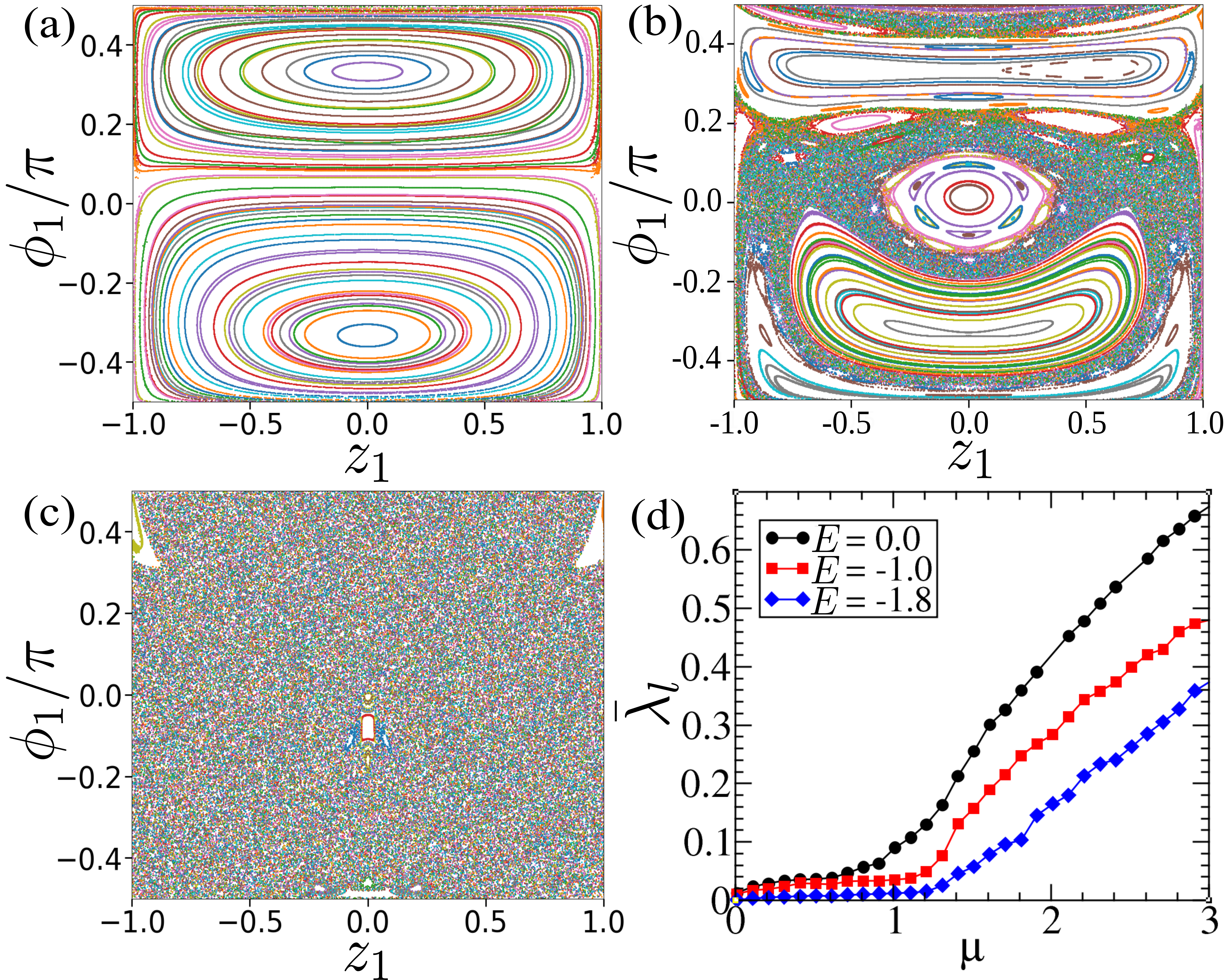}
	\caption{Poincar\'{e} sections computed for $z_2=0$ plane at a fixed energy $E=-1.0$ for increasing coupling strength, (a) $\mu=0.5$, (b) $\mu=1.2$, (c) $\mu=1.85$, depicting the onset of chaos. (d) Average Lyapunov exponent $\bar{\lambda}_l$ with increasing $\mu$ for different energies $E$, computed over an ensemble of 200 initial phase space points.}
	\label{poincare-section}
\end{figure}
%%%%%%%%%%%%%%%%%%%%%%%%%%%%%%%%%%%%%%%%%%%%%%%%%%%%%%%%%%%%%%%%%%%%%%
To capture the onset of chaos in the CT model, we study the Poincar\'{e} section \cite{Strogartz} at the $z_2=0$ plane for the dynamics with fixed energy $E$. As seen from Fig.\ref{poincare-section}(a-c), for small coupling strength $\mu<\mu_c$, the Poincar\'{e} section contains only periodic orbits, signifying regular motion in phase space. Increasing  $\mu$ leads to the emergence of a number of chaotic trajectories that coexist with the periodic orbits, manifesting an interesting mixed phase space behavior. For large values of $\mu$, the phase space is completely filled up with chaotic trajectories, and the system becomes fully chaotic. In order to quantify the local chaotic behavior, we compute the Lyapunov exponent (LE) $\lambda$ \cite{Strogartz,monodromy1}, which is a standard measure to quantify the degree of chaos. Typically, for a chaotic trajectory, a small initial perturbation $\delta \mathbf{X}(t=0)$ at the phase space point $\mathbf{X}=\{z_1,\phi_1,z_2,\phi_2\}$ grows exponentially in time $||\delta\mathbf{X}(t)||=e^{\lambda t} \,||\delta\mathbf{X}(0)||$, which yields the Lyapunov exponent,
\begin{eqnarray}
\lambda&=&\lim_{t \to \infty}\frac{1}{t}\ln \left(\frac{||\delta\mathbf{X}(t)||}{||\delta\mathbf{X}(0)||}\right)
\label{LE}
\end{eqnarray}
\noindent When the limit in Eq.(\ref{LE}) exists and $\lambda$ is positive, the trajectories are sensitive to the initial condition and thus become chaotic in nature. The Lyapunov exponent $\lambda$ can be computed by following the standard procedure described in \cite{monodromy1,monodromy2}. The overall chaotic behavior can be quantified by the average Lyapunov exponent $\bar{\lambda}_l$, which is obtained by averaging $\lambda_l$, computed for large sample of initial phase space points, corresponding to a fixed energy $E$. The growth of $\bar{\lambda}_l$ above the critical coupling $\mu_c$, captures the onset of chaos, as depicted in Fig.\ref{poincare-section}(d). Interestingly, the average LE $\bar{\lambda}_l$ exhibits an energy dependence across the energy band, which increases towards the centre of the band with $E=0$.
\subsection{Spectral statistics}
\label{Spectral_Statistics}
Although, it is not possible to quantify chaos from Lyapunov exponent quantum mechanically, due to the absence of phase space trajectories, however its signature can be captured from spectral statistics of the corresponding Hamiltonian.  
According to Berry Tabor’s conjecture \cite{Berry_Tabor}, the energy level spacing distribution of the  system with regular dynamics follows Poisson statistics, whereas Bohigas-Giannoni-Schmit (BGS) conjecture \cite{BGS} suggests, Wigner-Dyson distribution of level spacing for a classically chaotic system. 
To analyze the spectral statistics of the energy eigenvalues, we diagonalize the Hamiltonian given in Eq.\eqref{Coupled Top} and obtain the eigenvalues $\mathcal{E}_{\nu}$ corresponding to eigenvectors $\ket{\psi_{\nu}}$. Since the spectral statistics is performed for a particular symmetry sector, it is important to find out the symmetries of the CT model. The Hamiltonian in Eq.\eqref{Coupled Top} remains invariant under the action of parity $\hat{\Pi} = e^{i\pi(\hat{S}_{1x} + \hat{S}_{2x})}$ and spin exchange $(S_1 \leftrightarrow S_2 )$ operator $\hat{\mathcal{O}}$ \cite{mondal}, which flips the indices of basis states $\ket{m_{1z},m_{2z}}$, where $m_{iz}$ are the quantum numbers of $\hat{S}_{iz}$. Both the operators $\hat{\Pi}, \hat{\mathcal{O}}$ posses two eigenvalues $\pm 1$, which we call as even (+1) and odd (-1). For spectral statistics, we  consider the symmetry sector, for which both the eigenvalue of $\hat{\Pi}$ and $\hat{\mathcal{O}}$ are +1, without any loss of generality.
%%%%%%%%%%%%%%%%%%%%%%%%%% Fig:5 Spectral analysis %%%%%%%%%%%%%%%%%%%%%%%%%%%%%%%%%
\begin{figure}
	\centering
	\includegraphics[height=3.9cm,width=8.65cm]{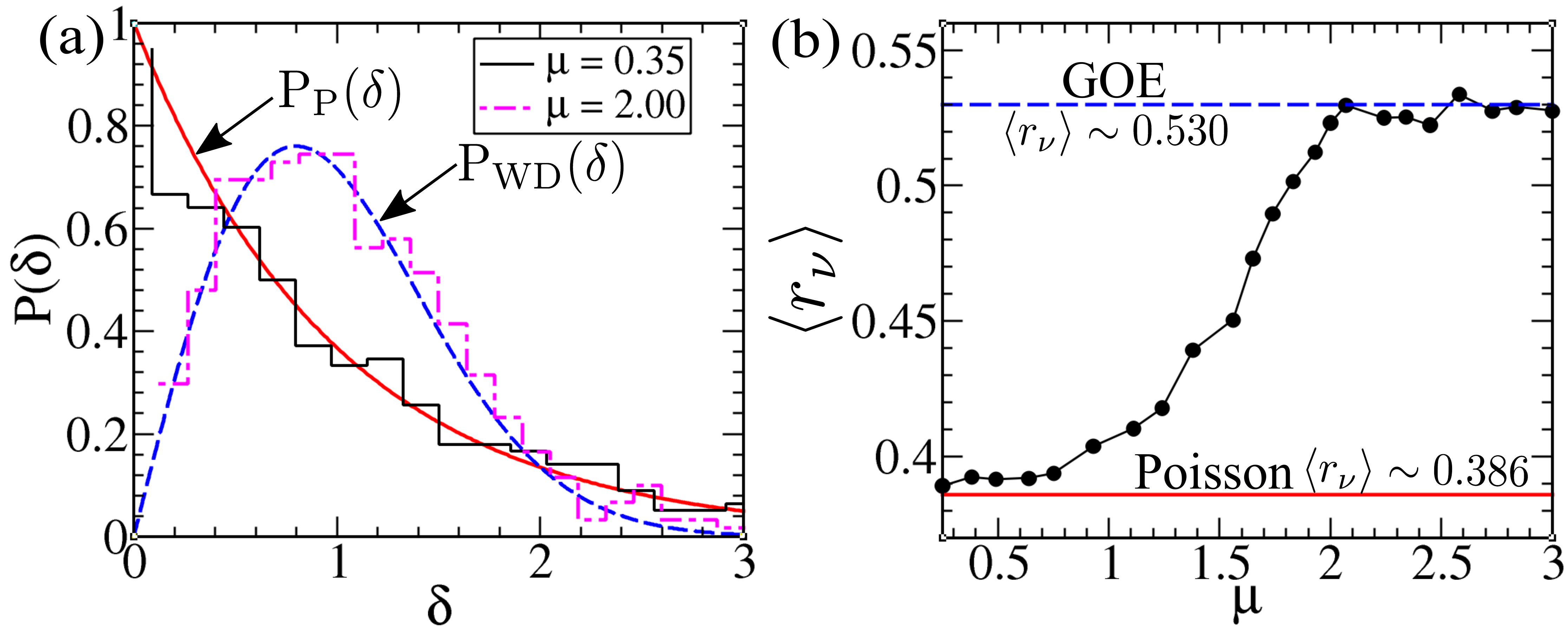}
	\caption{ (a) Level spacing distribution for regular ($\mu=0.35<\mu_c$) and chaotic ($\mu=2.00>\mu_c$) regime, which are compared with $\text{P}_{\text{P}}(\delta)$ and $\text{P}_{\text{WD}}(\delta)$ respectively. (b) Variation of average ratio of consecutive level spacing  $\langle r_{\nu} \rangle$ with increasing coupling strength $\mu$. Solid red (dashed blue) line  indicates the value of $\langle r_{\nu} \rangle$ corresponding to Poisson and GOE statistics respectively.}
	\label{level spacing}
\end{figure}
%%%%%%%%%%%%%%%%%%%%%%%%%%%%%%%%%%%%%%%%%%%%%%%%%%%%%%%%%%%%%%%%%%%%%%%%%%%%%%%%%%%%%%%
Next, we arrange the eigenvalues corresponding to the said symmetry sector, in ascending order and compute the consecutive level spacings, $\delta_{\nu} = \mathcal{E}_{\nu +1} -\mathcal{E}_{\nu}$. Then, the normalized level spacing distribution is obtained by keeping the mean to be unity, following the standard procedure \cite{Haake,Mehta}.
As seen from Fig.\ref{level spacing}(a), the obtained level spacing distribution of energy eigenvalues agrees well with the  Poisson statistics, $\text{P}_{\text{P}}(\delta) = \exp(-\delta)$ for small coupling strength $\mu<\mu_c$, on the other hand, the spacing distribution exhibits level repulsion and approaches the  Wigner-Dyson (WD) distribution, $\text{P}_{\text{WD}}(\delta) = \frac{\pi}{2}\delta\exp(-\frac{\pi}{4}\delta^2)$ corresponding to random matrix theory (RMT) of Gaussian orthogonal ensemble (GOE) for an intermediate range of coupling strength above $\mu_c$, where the underlying phase space becomes fully chaotic. 
The crossover from Poisson to  Wigner-Dyson statistics of spacing distribution can also be captured from the average ratio of consecutive level spacing  $\langle r_{\nu} \rangle$ \cite{Bogomonly_2}, which is given by ,
\begin{eqnarray}
\langle r_{\nu} \rangle &=& \left\langle \frac{\text{min}(\delta_{\nu} , \delta_{\nu+1} )}{\text{max}(\delta_{\nu} , \delta_{\nu+1} )}\right\rangle
\end{eqnarray}
For Poisson matrix of level spacing,
$\langle r_{\nu} \rangle \approx 0.386$ \cite{Bogomonly_2}, whereas in case of Gaussian orthogonal ensemble (GOE) of RMT, $\langle r_{\nu} \rangle \approx 0.53 $ \cite{Bogomonly_2}. Fig.\ref{level spacing}(b) shows the variation of $\langle r_{\nu} \rangle $ with $\mu$, which clearly exhibits the crossover from Poisson to the WD statistics.
\section{Ergodic behavior and multifractality of eigenstates}
\label{Ergodic}
\begin{figure}
	\centering
	\includegraphics[height=3.89cm,width=8.67cm]{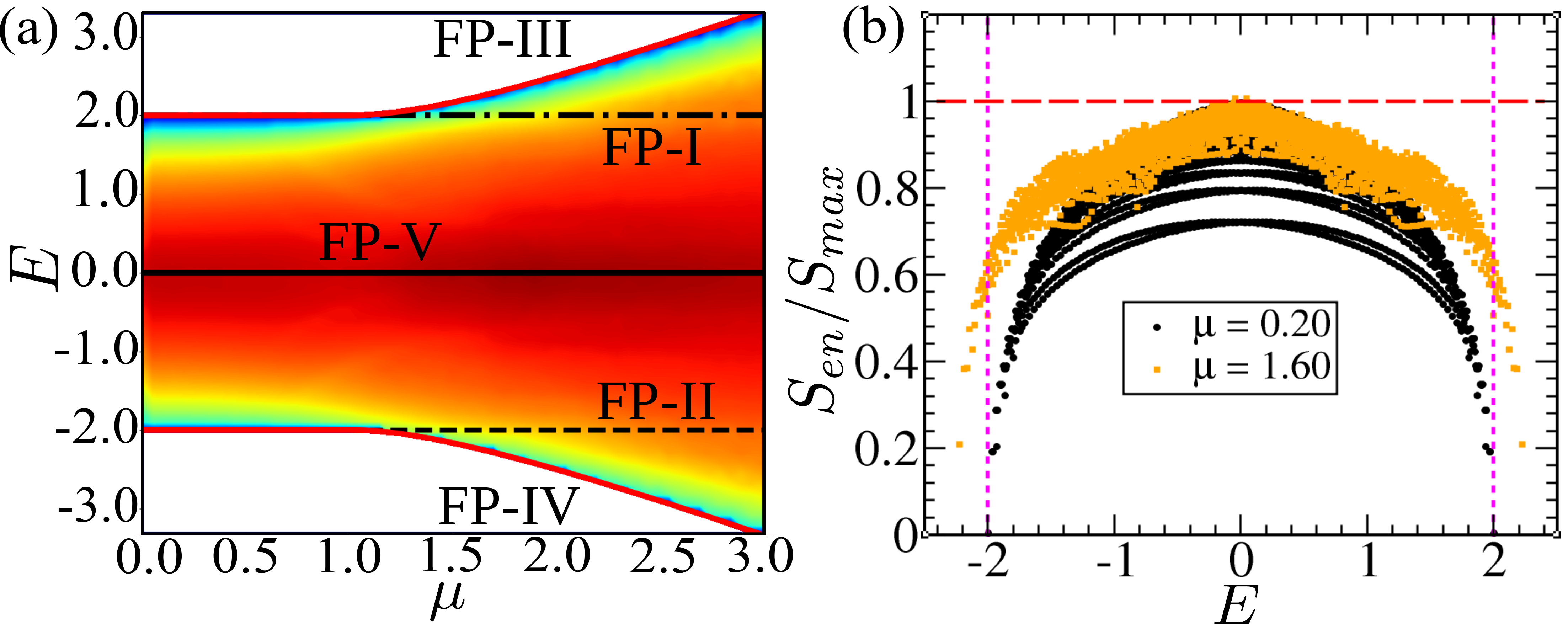}
	\caption{ Ergodic behavior of coupled top model: (a)Variation of relative entanglement entropy (EE) $S_{en}/S_{max}$ (color plot) of eigenstates with energy densities $E$ for increasing coupling strength $\mu$. (b) Variation of relative EE  of eigenstates with energy density $E$ for different coupling $\mu$. Horizontal red dashed line represents the maximum EE at unity. Pink dashed lines represent energy densities $E_c$ corresponding to ESQPT.}
	\label{Entanglement}
\end{figure}

In the previous section, we have investigated the onset of chaos in CT model at an intermediate coupling  above the critical point, both classically and at the quantum level from the spectral statistics. In this section, we present a more detailed study of the eigenstates, which can reveal interesting phenomenon related to ergodicity and non-equilibrium dynamics. Since the entropy is related to the degree of ergodicity, we study entanglement entropy (EE) of the eigenstates $\ket{\psi_{\nu}}$ of the Hamiltonian given in Eq.\eqref{Coupled Top}. To obtain the EE, we compute the reduced density matrix $\hat{\rho}_S$ of a particular spin sector $S$, as follows,
\begin{eqnarray}
\hat{\rho}_S &=& \text{Tr}_{\bar{S}}\ket{\psi_{\nu}}\bra{\psi_{\nu}}
\end{eqnarray}
where $\text{Tr}_{\bar{S}}$ denotes the partial trace with respect to the other spin sector. The EE quantifies the degree of entanglement, which is given by,
\begin{eqnarray}
S_{en}&=& -\text{Tr}\,\hat{\rho}_S\,\log \hat{\rho}_S
\label{Entanglement_entropy}
\end{eqnarray}
We quantify the degree of ergodicity of a state by comparing the EE with its maximum value,
\begin{eqnarray}
S_{\text{max}} &\simeq & \text{log}(\mathcal{D}_A)-\mathcal{D}_A/2\mathcal{D}_B
\label{max_EE}
\end{eqnarray}
which is obtained for a random state, partitioned into subsystems A(B) with dimensions $\mathcal{D}_A(\mathcal{D}_B )$ \cite{Page}. Therefore, the relative EE $S_{en}/S_{max}$ attains a maximum value at unity. In CT model, we have $\mathcal{D}_A= \mathcal{D}_B=2S+1$, hence the maximum value of EE is given by $S_{\text{max}} \simeq  \text{log}(2S+1)-1/2$. From the comparison of the Lyapunov exponent obtained for the edge and centre of the energy band (shown in Fig.\ref{poincare-section}(d)), we observe a variation in degree of chaos across the band, which is maximum at the band centre, with classical energy $E=0$.
Interestingly, at the quantum level, the degree of ergodicity in terms of the relative EE $S_{en}/S_{max}$  reveals the similar behavior with increasing energy density $E$ across the energy band which is peaked at the centre with energy density $E=0$ (see Fig.\ref{Entanglement}(a)). Although, in general, the notion of chaos is not quite the same as ergodicity in quantum systems.
%%%%%%%%%%%%%%%%% Fig:6 ergodic behaviour in terms of EE %%%%%%%%%%%%%%%%%%%%%%%%%%%
%%%%%%%%%%%%%%%%%%%%%%%%%%%%%%%%%%%%%%%%%%%%%%%%%%%%%%%%%%%%%%%%%%%%%%
Overall ergodic behavior of the CT model, in terms of the relative EE with increasing coupling $\mu$, across the QPT is shown as color scale plot in Fig.\ref{Entanglement}(a). It is evident from Fig.\ref{Entanglement}  that, after QPT, the symmetry broken eigenstates near the band edge are less ergodic compared to the symmetry unbroken states, which are separated by the ESQPT lines (FP-I,II). Above QPT, the relative EE of the eigenstates near band centre with $E\approx 0$ increases and attains its maximum value $S_{en}/S_{max}\approx 1$  at an intermediate coupling strength $\mu \approx 1.85$ (see Fig.\ref{Entanglement}(b)), where the system becomes fully chaotic and $\langle r_{\nu} \rangle$ approaches to GOE limit. As discussed earlier, since the model exhibits integrable behavior in the two extreme limits of coupling ($\mu \ll 1$ and $\mu \gg 1$), the system becomes maximally ergodic within a  moderate range of coupling across $\mu\approx 1.85$, which is the main focus in the later part of this work.
\subsection{Multifractality of eigenstates}
\label{multifractality}
%%%%%%%%%%%%%%%%% Fig:7 Multifractal dimension %%%%%%%%%%%%%%%%%%%%%%%%%%%%%
\begin{figure}
	\centering
	\includegraphics[height=8cm,width=8.7cm]{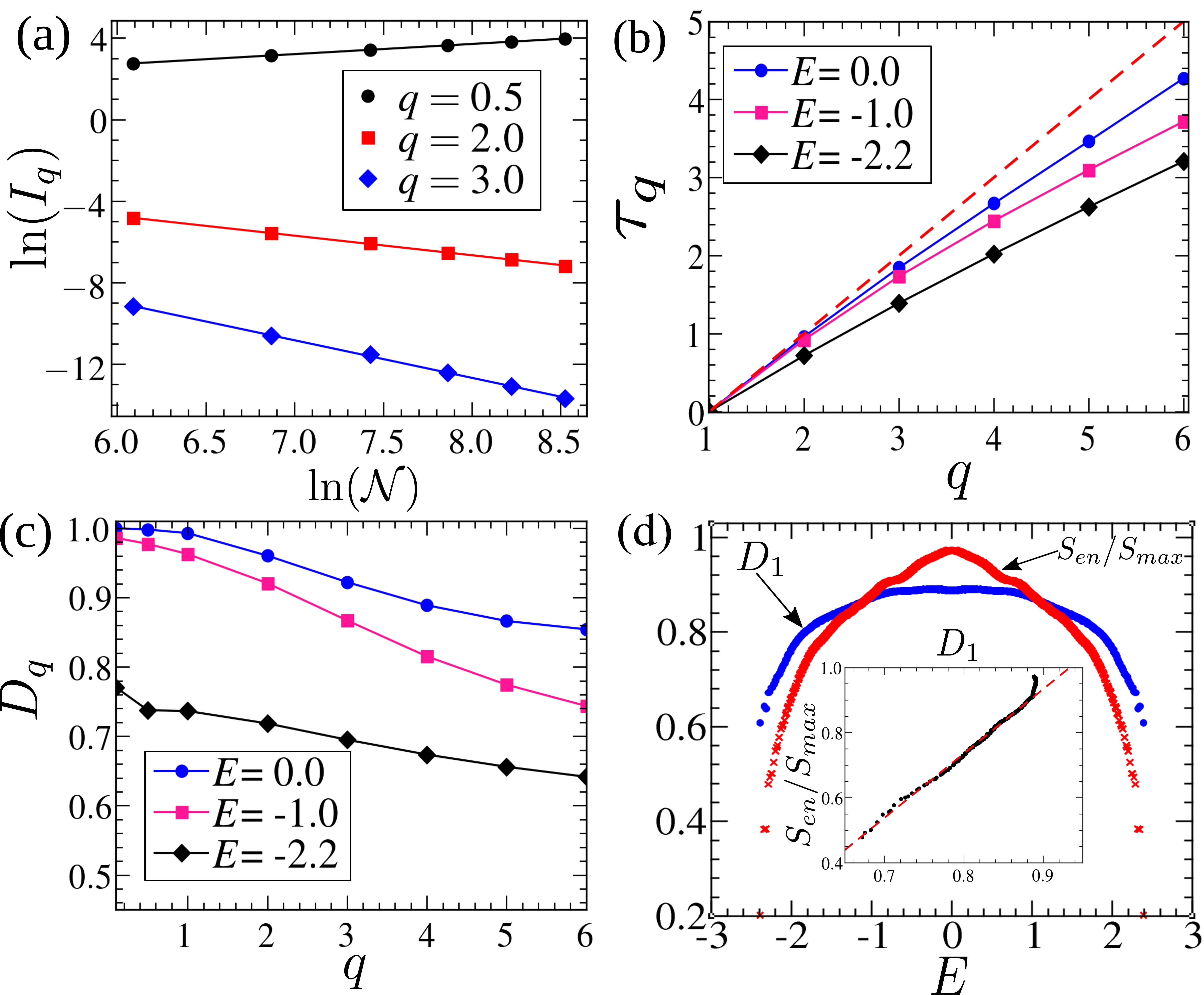}
	\caption{Multifractality of the energy eigenstates: (a) Scaling of generalized IPR $I_{q}$ with dimension $\mathcal{N}$ for eigenstates at the band centre with energy density $E\approx0$.  The solid lines represent the linear fit. (b) Variation of $\tau_q$ with $q$ for eigenstates with different values of $E$. The red dashed line represents  $\tau_q=q-1$ with $D_q=1$. (c) Multifractal dimension $D_q$ (obtained from $\tau_q$) as a function of $q$ for different energy densities $E$.  (d) Comparison between the multifractal dimension $D_1$ and relative EE $S_{en}/S_{max}$, for eigenstates with increasing energy density $E$ across the energy band. The linear behavior of $D_1$ with $S_{en}/S_{max}$ curve is shown in the inset of (d) and the red dashed line represents the corresponding linear fit. For all the figures, $\mu = 1.85$.}
	\label{Dq}
\end{figure}
%%%%%%%%%%%%%%%%%%%%%%%%%%%%%%%%%%%%%%%%%%%%%%%%%%%%%%%%%%%%%%%%%%%%%%
The localization and ergodic behavior of a state $\ket{\psi}=\sum_{i=1}^{\mathcal{N}} \psi^i \ket{i}$ can be determined from the participation of the computational basis $\ket{i}$ in the formation of said state.  For a localized state, a few basis states participate with non vanishing $\psi^i$, whereas, for a fully ergodic state $\psi^i \sim 1/\mathcal{N}$, where $\mathcal{N}$ is the dimension of the Hilbert space. In addition, the multifractal state lies in between these two extremes and their presence in a quantum many body system can give rise to an interesting non-ergodic behavior \cite{Bogomolny_multifractality,Lea_Santos_multifractality}. Recently, it has been observed that the ground state of the quantum many body systems exhibit multifractality and the appearance of QPT in such systems can be captured from the change of multifractal dimensions \cite{Bogomolny_multifractality,Rodriguez}. In appendix \ref{multifractal_dimension_QPT}, we also discuss how the QPT in CT model can be captured from the multifractal dimension of the ground state computed in a suitably chosen basis.
The multifractal nature of the eigenstates $\ket{\psi_{\nu}}$ can be captured from the statistical analysis of the coefficients $\psi^{i}_{\nu}$  and their scaling with the dimensionality $\mathcal{N}$. To quantify the degree of localization of a state, the generalized  inverse participation ratio (IPR) is given by, 
\begin{eqnarray}
I_q=\sum_{i}|\psi^i|^{2q}
\label{IPR}
\end{eqnarray}
which reduces to the usual IPR for $q=2$. In general, $I_q\sim 1/\mathcal{N}^{\tau_q}$ for sufficiently large dimension $\mathcal{N}$. In case of a localized state, $\tau_q$ vanishes and $I_{q}$ becomes independent of $\mathcal{N}$, whereas for an extended ergodic state, $\tau_q\sim (q-1)$ as $|\psi^{i}|^2 \sim 1/\mathcal{N}$. Generally, the exponent can be written as $\tau_q = (q-1)D_q$, where $D_{q}$ is the $q$th multifractal exponent characterizing the multifractality of the state. Alternatively, $D_q$ can be calculated from the R\'{e}nyi entropy $S_R(q,\mathcal{N})$ as follows,
\begin{eqnarray}
D_q=\lim_{\mathcal{N} \to \infty}\frac{S_R(q,\mathcal{N})}{\ln\mathcal{N}}
\end{eqnarray}
where the R\'{e}nyi entropy is  defined by, 
\begin{eqnarray}
S_R(q,\mathcal{N})=-\frac{1}{(q-1)}\ln\left(\sum_{i}^{\mathcal{N}}|\psi^i|^{2q}\right).
\end{eqnarray}
For $q\rightarrow1$, the R\'{e}nyi entropy reduces to Shannon entropy $S_{\rm Sh} =-\sum_i^{\mathcal{N}} \big{\vert}\psi^i\big{\vert}^{2}\ln \big{\vert}\psi^i\big{\vert}^{2}$ which yields the exponent $D_{1}$, 
\begin{eqnarray}
D_1&=&\lim_{\mathcal{N} \to \infty}\frac{S_{\rm Sh}}{\ln\mathcal{N}}.
\end{eqnarray}
Note that, the numerical computations are performed in the $\hat{S}_{z}$ basis.

When the CT model becomes maximally chaotic ($\mu\approx$ 1.85), we investigate the multifractal properties of eigenstates with different energy densities using the above mentioned framework. For the eigenstates at the band centre ($E\approx 0$), the variation of generalized IPR $I_q$ with dimension $\mathcal{N}$ as shown in the log-log scale reveals the linear behavior (see Fig.\ref{Dq}(a)) and the slope yields the exponent $\tau_{q}$ for different $q$. As seen from Fig.\ref{Dq}(b), the behavior of $\tau_{q}$ with $q$ deviates from linearity, indicating multifractal nature of the eigenstates. The multifractal dimension $D_{q}$ can be extracted from $\tau_{q}$, and its variation with $q$ for different energy densities is shown in Fig.\ref{Dq}(c), which clearly exhibits the multifractal character of the eigenstates at different energy densities.
The eigenstates at the band centre with $E\approx 0$ have multifractal dimension close to unity, which indicates that these states approach the ergodic limit. In the numerical computations, the presence of finite size effect leads to larger deviation in $D_{q}$ for higher values of $q$, which is also present for eigenvectors in random matrices with finite dimensionality \cite{Haque}. Due to such finite size effects, it is difficult to conclude whether the states at band centre ($E\approx 0$) becomes fully ergodic or remain weakly ergodic. Away from the band centre, the multifractal dimension $D_{q}$ of the eigenstates decreases with increasing energy density $|E|$, indicating their non-ergodic nature. We plot the averaged multifractal dimension $D_{1}$ with increasing energy density $E$ across the energy band, which resembles with the behavior of relative EE, as depicted in Fig.\ref{Dq}(d). Interestingly, the linear variation of relative EE $S_{en}/S_{max}$ with  multifractal exponent $D_{1}$ (see inset of Fig.\ref{Dq}(d)) has recently been observed for non-ergodic `sparse random pure states' discussed in \cite{Khaymovich}. This behavior indicates that, both multifractal dimension $D_{1}$ and relative EE $S_{en}/S_{max}$ can quantify the degree of ergodicity, which is also reflected in the non-equilibrium dynamics, as discussed in \ref{non_ergodic_dynamics}.

\section{Quantum dynamics}
\label{quantum_dynamics}
In this section, we study the non-equilibrium dynamics of coupled top model to investigate the onset of chaos and related ergodic behavior. The time evolution of various observables can be studied from the time evolved state $\ket{\psi(t)} = e^{-\imath\hat{\mathcal{H}}t}\ket{\psi(0)}$ for a suitably chosen initial state $\ket{\psi(0)}$. To compare the quantum dynamics with its classical counterpart, we use the prescription of spin coherent state \cite{Radcliffe},
\begin{eqnarray}
\ket{z,\phi}=\left(\frac{1+z}{2}\right)^{S}\,\exp{\sqrt{\frac{1-z}{1+z}}e^{i\phi} \hat{S}_-}\ket{S,S}
\end{eqnarray}
where $z = \cos\theta$, with $\theta$ and $\phi$ representing the orientation of the corresponding classical spin vector. The two spins of the CT model can be represented as the product of two corresponding coherent states,
\begin{eqnarray}
\ket{\psi_c}=\ket{z_1,\phi_1,z_2,\phi_2}&=&\ket{z_1,\phi_1} \otimes \ket{z_2,\phi_2}
\end{eqnarray}
which provides a semiclassical description of the phase space point $\{z_{1},\phi_{1},z_{2},\phi_{2}\}$. Such choice of an initial state as the spin coherent state allows us to probe the local ergodic behavior. Following this prescription, we next study the onset of chaos and the ergodic behavior from the non-equilibrium dynamics.

\subsection{Classical quantum correspondence}
\label{classical_quantum_correspondence}
%%%%%%%%%%%%%%%%%%%%%% Fig:8 Classical variable vs expectation of S1z%%%%%%%%%%%%%%%
\begin{figure}
	\centering
	\includegraphics[height=7cm,width=8.5cm]{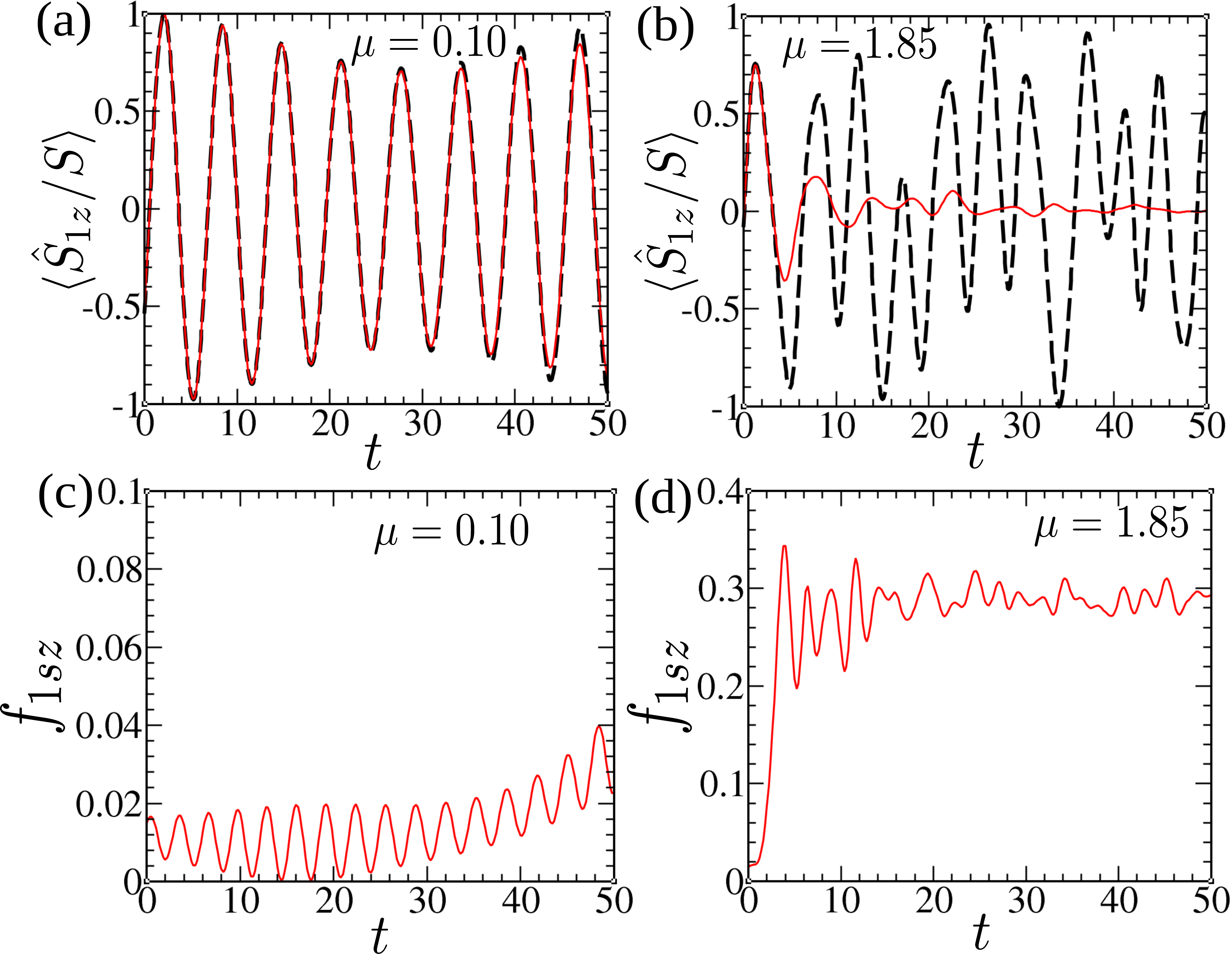}
	\caption{Time evolution of the observable $\langle \hat{S}_{1z} /S \rangle$ for (a) $\mu = 0.1$ and (b) $\mu = 1.85$. The time evolution obtained from quantum dynamics (solid red line) is compared with the results obtained from classical dynamics using Eq.\eqref{EOM} (black dashed line). Time evolution of the spin fluctuation $f_{1sz}=(\langle\hat{S}_{1z}^2\rangle-\langle\hat{S}_{1z}\rangle^2)/S^2$ for (c)  $\mu=0.1$ and (d) $\mu=1.85$.}
	\label{expectation of S1z}
\end{figure}
%%%%%%%%%%%%%%%%%%%%%%%%%%%%%%%%%%%%%%%%%%%%%%%%%%%%%%%%%%%%%%%%%%%%%%%%%%%%%%%%%%%%%%%%%%%% 
In this section, we compute the time evolution of the appropriately scaled physical observables obtained quantum mechanically and compare them with corresponding classical dynamics by solving the EOM, given in Eq.\eqref{EOM}. For this purpose, we choose the initial coherent state representing the initial classical phase space point. For small coupling regime $\mu \lesssim \mu_c$, when the system exhibits regular dynamics, the time evolution of the expectation value $\langle \hat{S}_{1z} /S \rangle$ is in good agreement with that of the classical dynamics \cite{Kicked_dickie}, as shown in Fig.\ref{expectation of S1z}(a). On the contrary, in the chaotic regime, the quantum evolution of the corresponding observable matches with its classical counterpart upto certain time scale, after which it deviates significantly with increasing time \cite{Kicked_dickie} (see Fig.\ref{expectation of S1z}(b)). Moreover, the fluctuation corresponding to the above mentioned observable in the chaotic regime (see Fig.\ref{expectation of S1z}(d)) is much larger compared to the regular regime (see Fig.\ref{expectation of S1z}(c)) \cite{Kicked_dickie} and, systematic study of such fluctuations allows us to quantify chaos, which is discussed in the next subsection.
\subsection{Manifestation of ergodicity in non-equilibrium dynamics }
\label{non_ergodic_dynamics}
To this end, we study the local ergodic behavior of CT model from the dynamics of different physical quantities. As seen from the analysis of the eigenstates in the previous section, the ergodic behavior of the eigenstates depends on their energy density $E$, even when the system becomes maximally chaotic.
%%%%%%%%%%%%%%%%%%% Fig:9 Classical variable vs expectation of S1xyz-Energy %%%%%%%%
\begin{figure}[H] 
	\centering
	\includegraphics[height=4.0cm,width=8.7cm]{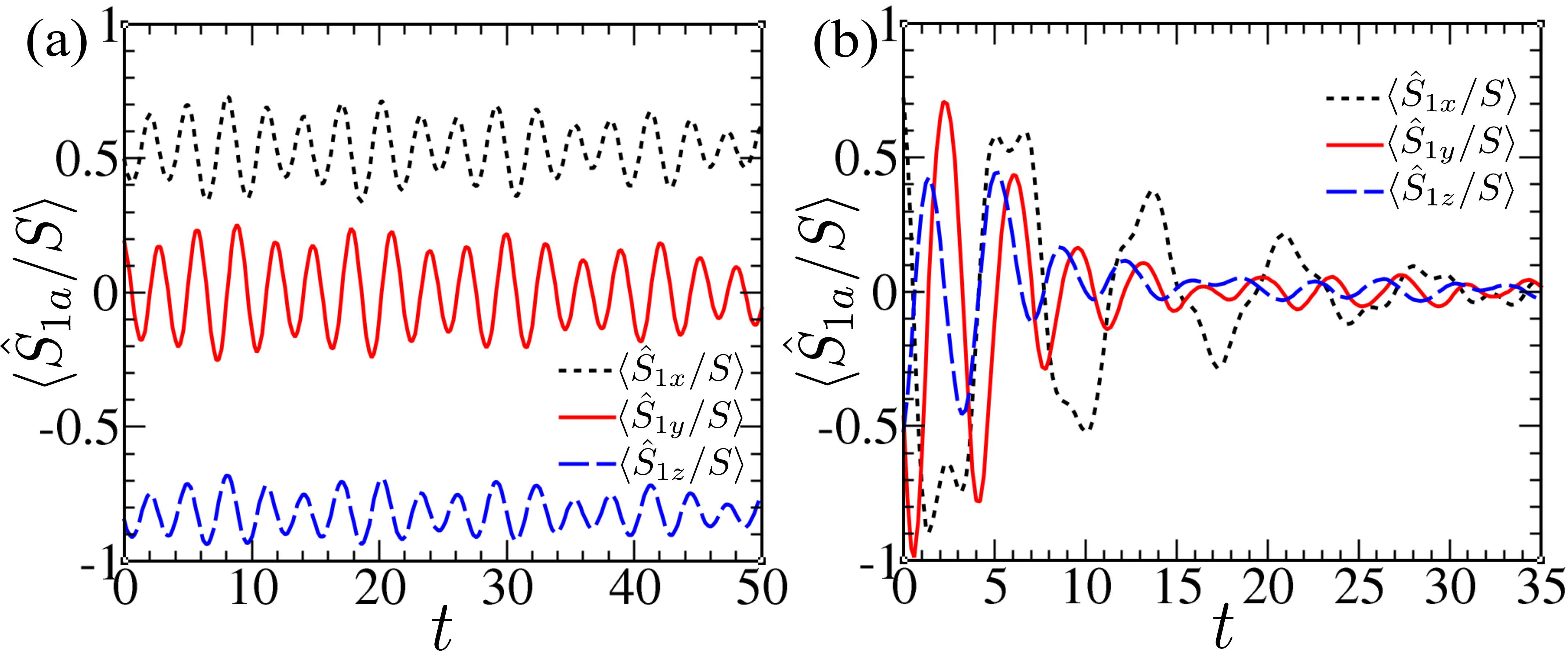}
	\caption{Dynamics of different components of the spins ($\langle \hat{S}_{1x,y,z} /S \rangle$) corresponding to the initial coherent state (a) near the band edge with $E = -2.3$ and (b) at the band centre for $E = -0.01$ . For both the figures, $\mu=1.85$.}
	\label{expectation of S1xyz-Energy}
\end{figure}
%%%%%%%%%%%%%%%%%%%%%%%%%%%%%%%%%%%%%%%%%%%%%%%%%%%%%%%%%%%%%%%%%%%%%%%%%%%%%%%%%%%%%%%%%%%%
%%%%%%%%%%%%%%%%%%%%%%%%% Fig:10 Erdodicity from dynamics %%%%%%%%%%%%%%%%%%%%%%%%%%
\begin{figure*}
	\centering
	\includegraphics[height=9.5cm,width=16.5cm]{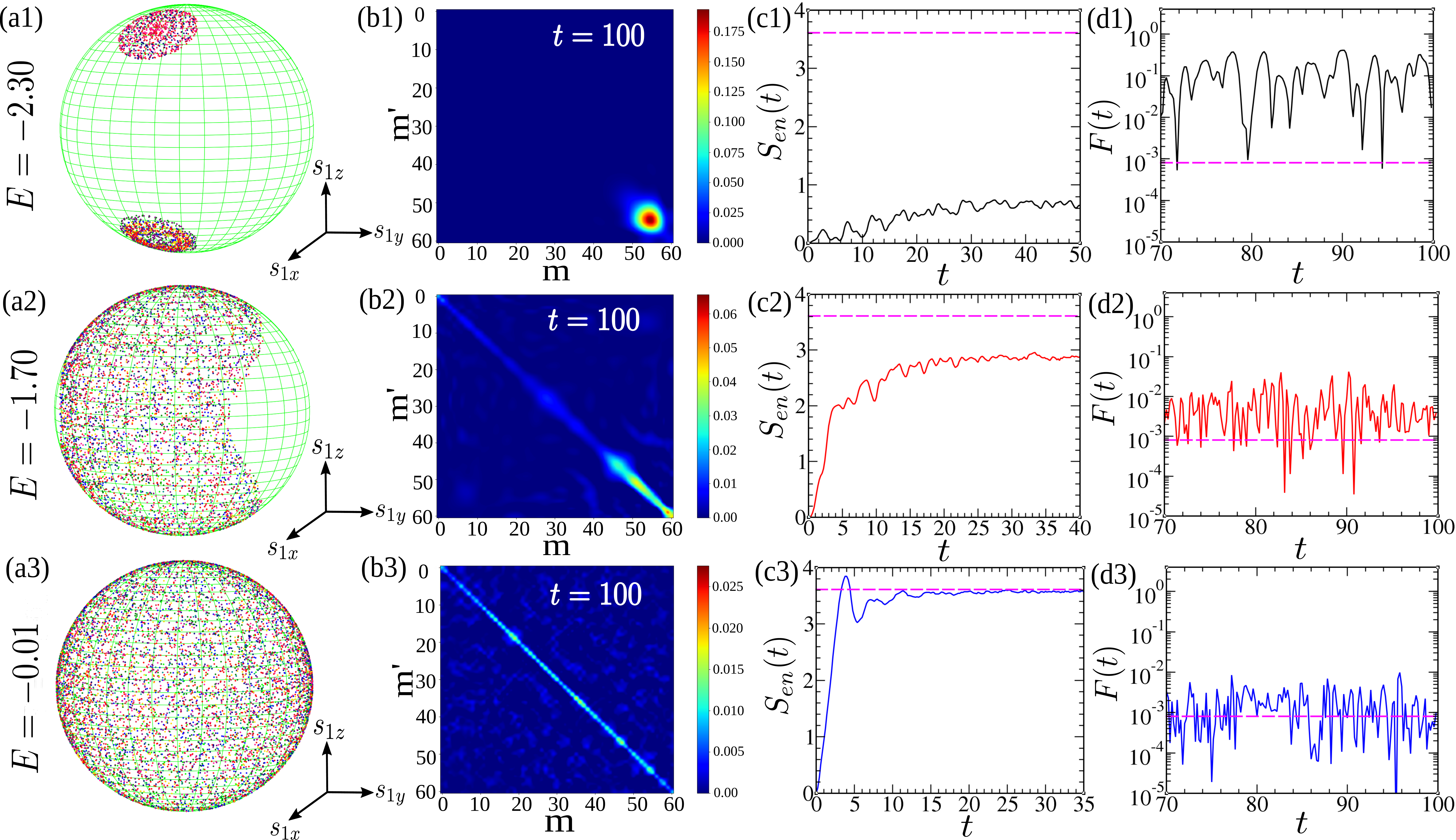}
	\caption{Manifestation of degree of ergodicity in non-equilibrium dynamics of different physical quantities, with increasing energy density $E$ (as shown in different rows). Column(a): The classical phase space trajectories on Bloch sphere with fixed classical energy $E$. Column(b): Color scale plot of the elements of final reduced density matrix $(\hat{\rho}_{\mathcal{S}})_{mm'}$ (in the basis of $\hat{S}_z$), after time evolution of an initial coherent state representing a phase space point corresponding to energy density $E$. Column(c) and (d) represent the time evolution of EE and survival probability respectively for the same initial coherent states. Horizontal pink dashed lines in column(c) and (d) denote maximum limit of EE (Eq.(\ref{max_EE})) and  GOE limit $3/\mathcal{N}$ of survival probability  respectively. Parameters chosen: $\mu=1.85$.}
	\label{ergodicity_q_vs_c}
\end{figure*}
%%%%%%%%%%%%%%%%%%%%%%%%%%%%%%%%%%%%%%%%%%%%%%%%%%%%%%%%%%%%%%%%%%%%%%%%%%%%%%%%% 
%The energy states at the centre of the band with $E\approx 0$ are more ergodic compared to the symmetry broken states at the edge of the band. 
We probe such behavior from non-equilibrium dynamics of the physical observables , starting from an initial coherent state with a given energy density $E$, since the energy remains conserved during the time evolution. In the maximally chaotic regime, we study the dynamics of the spin components $\langle \hat{S}_{1a}/S \rangle$ $(a=x,y,z)$ by choosing two initial coherent states with energy density corresponding to the edge and centre of the energy band. For dynamics with energy close to the band edge, the expectation values of the spin components exhibit oscillation without reaching any steady state, as shown in Fig.\ref{expectation of S1xyz-Energy}(a). Moreover, the non vanishing value of $\langle \hat{S}_{1z}/S \rangle$ reflects the  symmetry broken nature of the states at the band edge. On the other hand, the initial state at the band centre ($E \approx 0$), evolves to a steady state with vanishing expectation value of all the spin components  (see Fig.\ref{expectation of S1xyz-Energy}(b)), which is signature of diagonal reduced density matrix corresponding to microcanonical ensemble (discussed later in this section), indicating the ergodic evolution. This dynamical behavior supports the fact that the states near the centre of the energy band approaches the ergodic limit, whereas the absence of steady states from the dynamics of an initial state at the band edge indicates non-ergodic evolution, which is consistent with our previous analysis discussed in Sec.\ref{Ergodic}. 

To understand the dynamical signature of degree of ergodicity across the energy band in a more definitive manner, we study the evolution of EE \cite{lewenstein,shohini_ghose,EE_expt} and the final reduced density matrix, as well the survival probability for different energy densities. 
For time evolution of $\ket{\psi(t)}$ starting from an initial state $\ket{\psi(0)}$, the survival probability is defined as,
\begin{eqnarray} 
F(t) = |\langle \psi(t) | \psi(0) \rangle|^2
\label{SP}
\end{eqnarray}
Physically, it describes the memory of the initial state, which decays from unity and approaches to the GOE limit $3/\mathcal{N}$ \cite{Izrailev_2,Lea_santos_Survival} for ergodic evolution. At long time, the deviation of survival probability from the GOE limit can quantify the degree of ergodicity. Similarly, in the ergodic evolution, it is expected that the reduced density matrix $(\hat{\rho}_{\mathcal{S}})_{mm'}=\bra{m}\hat{\rho}_{\mathcal{S}}\ket{m'}$ ($\ket{m},\ket{m'}$ are basis states of $\hat{S}_z$) approaches to the microcanonical form with non vanishing diagonal elements giving rise to maximum EE $S_{\rm max}$ given in Eq.\eqref{max_EE}. To gain more insight, we also study the corresponding phase space dynamics with fixed energy, projected on the Bloch sphere. The degree of ergodicity across the energy band from its edge to the centre is summarized in Fig.\ref{ergodicity_q_vs_c}. For classical energy close to the band edge below the critical energy of ESQPT, the classical trajectories are confined within a small region around the symmetry broken fixed points (FP-III), clearly exhibiting a localized structure in Bloch sphere, as shown in Fig.\ref{ergodicity_q_vs_c}(a1). With increasing energy density $E$, the diffusive behavior of the trajectories is observed, which occupy the Bloch sphere partially,  as shown in Fig.\ref{ergodicity_q_vs_c}(a2)), and eventually cover the full Bloch sphere for $E\approx 0$ (see Fig.\ref{ergodicity_q_vs_c}(a3)) corresponding to the centre of energy band. Such behavior of increasing degree of ergodicity across the energy band is also reflected from the final form of reduced density matrix, as well from the long time saturation of EE and survival probability, as depicted in respective columns of Fig.\ref{ergodicity_q_vs_c}(b,c,d). The degree of ergodicity of eigenstates across the energy band, as shown in Fig.\ref{Entanglement}, is also reflected in the non-equilibrium dynamics. Moreover, the classical dynamics provides a connection between the phase space mixing and energy dependent ergodic behavior of its quantum counterpart. In the next subsection, we discuss a newly developed technique to probe the degree of ergodicity in a more systematic manner.
\subsection{OTOC dynamics}
\label{OTOC dynamics}
\begin{figure}
	\centering
	\includegraphics[height=3.85cm,width=8.7cm]{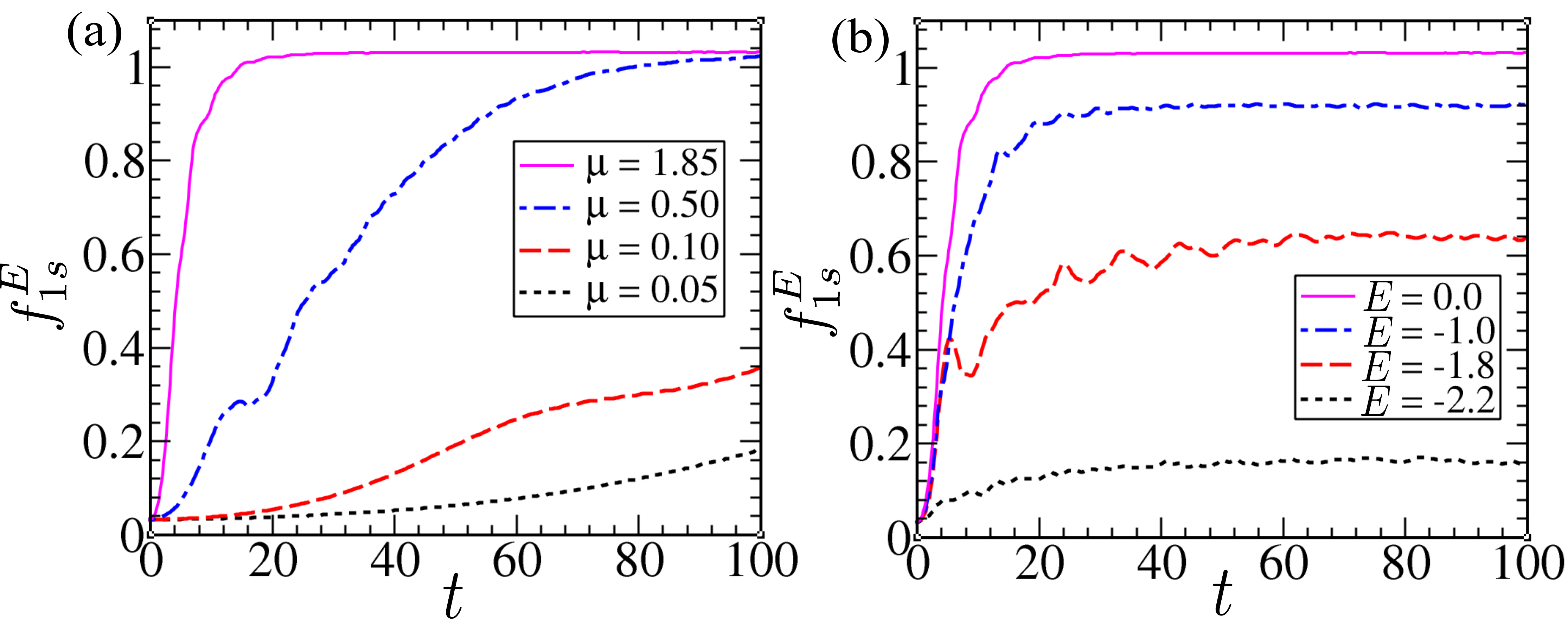}
	\caption{Dynamics of average FOTOC $f_{1s}^E$, computed for an ensemble of initial coherent states with fixed energy density $E$. (a) Time evolution of $f_{1s}^E$ for $E\approx0$, corresponding to the band centre, reflecting its change in degree of ergodicity with increasing coupling strength $\mu$. (b) Variation in dynamics of $f^E_{1s}$ for different energy densities $E$ across the energy band, reflecting the  maximum degree of ergodicity at the band centre for $\mu=1.85$.}	
	\label{FOTOC_and_chaos}
\end{figure}

In recent studies, it has been shown that the dynamics of `out-of-time-order correlator' (OTOC) can be used as a tool to detect  dynamical instability that can lead to chaos in an interacting quantum system \cite{stanford2,maldacena,K_hashimoto,Rozenbaum,Garttner1,butterfly_effect,Swingle1,
NMR,Trapped_ion,Fazio_otoc,A_M_rey2,Santos_otoc,
OTOC_instability_LMG,Garcia_mata,lakshminarayan,sray1,sudip_otoc}, which has also been implemented in NMR \cite{NMR} and trapped ion experiments \cite{Trapped_ion}. This was originally introduced in the context of superconductivity \cite{Larkin} and in recent years, it has also been applied in different areas of physics, most popularly in the context of black hole thermalization \cite{stanford2,maldacena}, and scrambling of quantum information \cite{Swingle1}. The growth rate of OTOC of appropriately chosen operators plays an analogous role of Lyapunov exponent for quantum systems \cite{Rozenbaum,A_M_rey2,Santos_otoc,OTOC_instability_LMG}, which can quantify the instability. Moreover, the recent studies reveal that its saturation value can also be used as a measure for degree of ergodicity \cite{Garcia_mata,lakshminarayan,sray1,sudip_otoc}. The OTOC of two operators $\hat{W}(t)$ and $\hat{V}(0)$ is defined as follows,
\begin{eqnarray}
O(t) = {\rm Tr}\hat{\rho}_0 \hat{W}^\dagger(t)\hat{V}^\dagger(0)\hat{W}(t)\hat{V}(0)
\end{eqnarray}
where $\hat{W}(t)$ represents the evolution of the operator at time $t$ and $\hat{\rho}_0$ is an appropriately chosen initial density matrix. This is  related to the unequal time commutator $C(t)$ defined as, 
\begin{eqnarray}
C(t) = {\rm Tr}\hat{\rho}_0[\hat{W}(t),\hat{V}(0)]^{\dagger}[\hat{W}(t),\hat{V}(0)]
\end{eqnarray}
which measures the non commutativity between these operators evolving with time, even if the operators commute initially. Both $C(t)$ and $O(t)$ are related by $C(t)=2(1-{\rm Re}O(t))$, for unitary $\hat{W}$ and $\hat{V}$. A generalization of OTOC known as the fidelity out-of-time-order correlator (FOTOC) is defined for $\hat{W}=e^{\imath\delta\phi\hat{G}}$ and $\hat{V}=\hat{\rho}_0$, for any hermitian operator $\hat{G}$ and $\hat{\rho}_0 =\ket{\psi(0)}\bra{\psi(0)}$, where $\ket{\psi(0)}$ is the initial state. For a sufficiently small perturbation $\delta\phi\ll 1$, the FOTOC is related to the fluctuation $f_{G}$ of the operator $\hat{G}$ \cite{A_M_rey2,Santos_otoc},
\begin{eqnarray}
1-\mathcal{F}_G = \left(\langle \hat{G}^2\rangle - \langle \hat{G} \rangle^2\right)\delta\phi^2\equiv f_{G} \delta\phi^2.
\label{FOTOC-definition}
\end{eqnarray} 
 Therefore, measuring the fluctuations of appropriately chosen operators can quantify the degree of ergodicity. In the present work, we study the FOTOC corresponding to the sum of fluctuations of all the  spin components in a particular spin sector, which is given by,
\begin{eqnarray}
f_{is} = \sum_{a=x,y,z}f_{isa}=\sum_{a=x,y,z} \left(\langle \hat{S}^2_{ia}\rangle - \langle \hat{S}_{ia} \rangle^2 \right)/S^2.
\label{FOTOC}
\end{eqnarray}
In order to probe the ergodicity in an energy resolved manner, we define energy dependent FOTOC of spins $f^{E}_{is}$, which is obtained by averaging $f_{is}$ over an ensemble of initial coherent states with a fixed energy density $E$. As depicted in Fig.\ref{FOTOC_and_chaos}(a), the transition to chaos with increasing coupling strength $\mu$ can be captured from the growth rate of $f^{E}_{is}$ with $E \approx 0$. Moreover, in the chaotic regime, its saturation value approaches to unity, which is consistent with the fact that the states at the band centre with $E \approx 0$ become ergodic. Near the maximally chaotic region at $\mu \approx 1.85$, we investigate the energy dependent degree of ergodicity from the dynamics of $f^E_{is}$ for different energy densities, which is shown in Fig.\ref{FOTOC_and_chaos}(b). With increasing energy density $E$, corresponding to the energy band, from its edge to the centre,  the growth rate of $f^E_{is}$ as well as its saturation value increases, which finally approaches to unity for $E=0$ (see Fig.\ref{FOTOC_and_chaos}(b)). Such behavior, observed in FOTOC dynamics confirms the variation of degree of ergodicity across the energy band shown in Fig.\ref{Entanglement}.
%%%%%%%%%%%%%%%%%%%%%&&&&&&&&&& Fig:11 FOTOC and chaos%%%%%%%%%%%%%%%%%%%%%%%%%%%%%
%%%%%%%%%%%%%%%%%%%%%%%%%%%%%%%%%%%%%%%%%%%%%%%%%%%%%%%%%%%%%%%%%%
The present analysis opens up the possibility for experimental investigation of local ergodic behavior of a quantum system using FOTOC technique, which has already been implemented in trapped ion system \cite{A_M_rey2}.
%%%%%%%%%%%%%%%%%%%%%%%%%% QUANTUM SCARS %%%%%%%%%%%%%%%%%%%%%%%%%%%%%%%%%%%%%%%%%%
\section{Formation of quantum scars}
\label{Quantum scar}
The investigation related to the fate of the unstable steady states in coupled top (CT) model reveals another source of deviation from ergodicity, due to the quantum scarring phenomena. In recent years, the study of quantum scars has regained interest due to the experimental observation of periodic revival and athermal behavior of some special choice of initial state \cite{Bernien}, which is attributed to many body quantum scar \cite{Turner,Motrunich,emergent_SU2,M_Lukin1,abanin,Silva_scar}.
In the CT model, two types of scars are observed, which appear as reminiscence of unstable fixed points (FPs) and periodic orbits, as mentioned in our previous work \cite{mondal}. In this section, we elaborate the discussion on above mentioned scars and  present new results related to detection of dynamical signature of scars using FOTOC technique. In addition, we identify new scars and provide a systematic statistical analysis to identify the scarred eigenstates.
\subsection{Scars of unstable fixed points }
\label{Scar_FP}
\begin{figure}
	\centering
	\includegraphics[height=11.65cm,width=9.0cm]{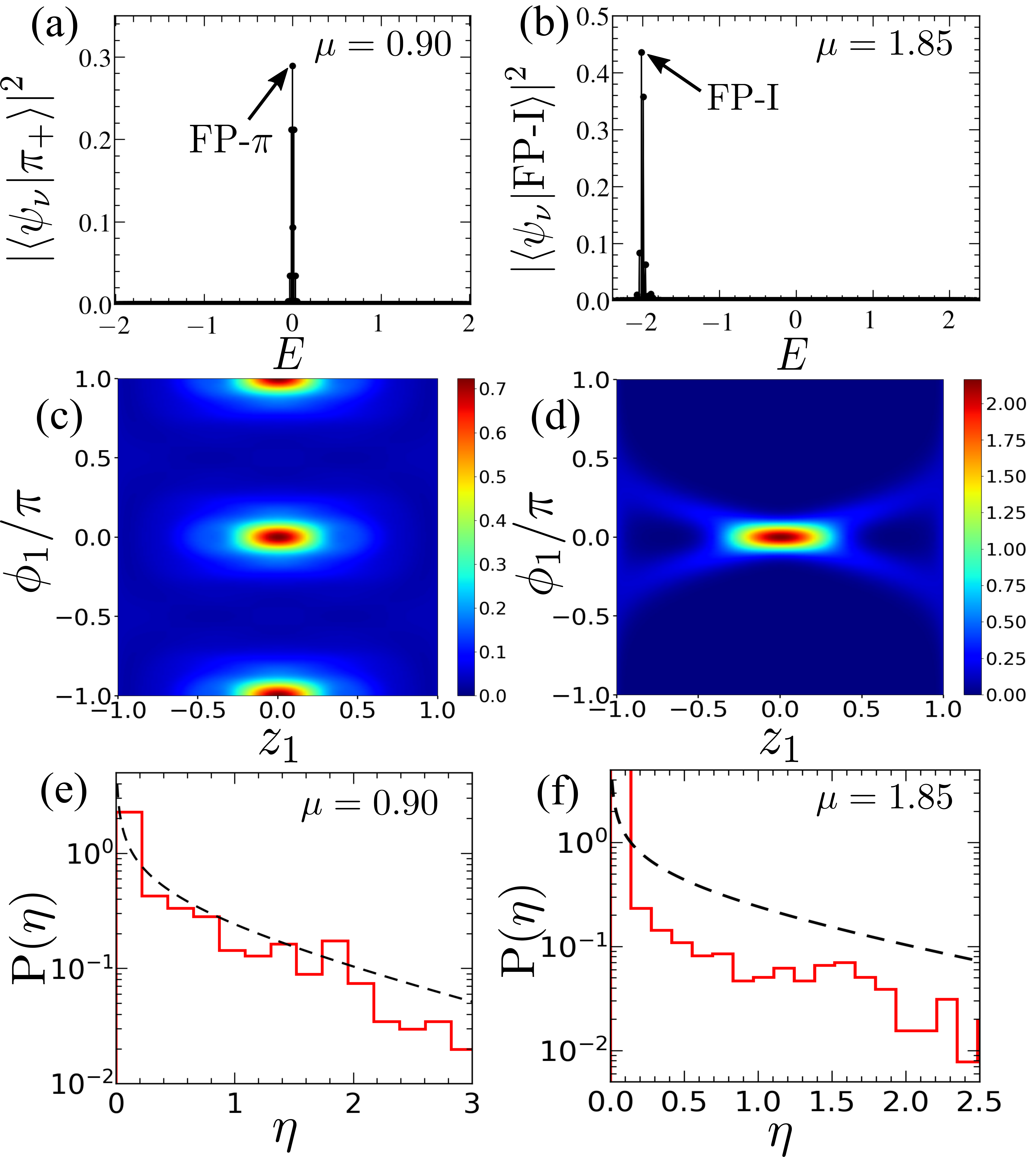}
	\caption{Identification of eigenstates bearing the scars of unstable FPs. Overlap $|\langle \psi_c|\psi_{\nu}\rangle|^2$ of eigenstates $\ket{\psi_{\nu}}$ with coherent states $\ket{\psi_c}$ corresponding to (a) the unstable $\pi-$mode and (b) FP-I. (c), (d) Husimi distribution for a particular spin sector of the eigenstate $\ket{\psi_{\nu}}$ having maximum overlap (marked by arrowhead in (a), (b)). (e), (f) Distribution $\text{P}(\eta)$ of the elements $\eta=|\psi_{\nu}^i|^2\mathcal{N}$ of the scarred eigenstates (marked by arrowhead in (a), (b)). The black dashed line represents the Porter-Thomas (PT) distribution.}
	\label{Scar-FP}
\end{figure}

To this end, we discuss about the quantum scars formed due to the unstable FPs. The reminiscence of the unstable FPs can be detected in some energy eigenstates in the form of a scar. Since such unstable FPs can be represented  semiclassically in terms of the coherent states $\ket{\psi_c}$, the corresponding scarred eigenstates $\ket{\psi_{\nu}}$ can be identified from the large overlap $|\langle \psi_c|\psi_{\nu}\rangle|^2\gg 1/\mathcal{N}$ \cite{abanin,sinha1}, where $\mathcal{N}=(2S+1)^2$ is the dimension of the Hilbert space. In the ergodic regime, such overlap becomes $O(1/\mathcal{N})$, indicating delocalization. To visualize the scars, we compute the Husimi distribution obtained from the reduced density matrix $\hat{\rho}_S$ of the corresponding scarred eigenstate $\ket{\psi_{\nu}}$, 
\begin{eqnarray}
Q(z,\phi)=\frac{1}{\pi}\bra{z,\phi}\hat{\rho}_\mathcal{S}\ket{z,\phi},
\label{Husimi}
\end{eqnarray} 
which provides a semiclassical phase space distribution. As seen from Fig.\ref{Scar-FP}(c,d), the Husimi distribution of the scarred eigenstates exhibits accumulation of phase space density near the corresponding unstable FPs, retaining their reminiscence at the quantum level. 

In the present case, above the critical coupling $\mu_c$, the unstable `$\pi$-mode' (FP-V), as well the two symmetry unbroken unstable steady states (FP-I, FP-II), as discussed in Sec.\ref{Classical}, can manifest in the form of scars in the energy eigenstates. Since there are two degenerate spin configurations corresponding to the $\pi$-mode, to describe such state semiclassically, we consider the following linear combination of the coherent states,
\begin{eqnarray}
\ket{\pi_+}=\frac{1}{\sqrt{2}}\left(\ket{0,0}\otimes \ket{0,\pi}+\ket{0,\pi}\otimes \ket{0,0}\right)
\label{pi_mode_eqn}
\end{eqnarray}
%%%%%%%%%%%%%%%%%%%%%%%%% Fig:12 Scar of fixed points %%%%%%%%%%%%%%%%%%%%%%%%%%%%%%
%%%%%%%%%%%%%%%%%%%%%%%%%%%%%%%%%%%%%%%%%%%%%%%%%%%%%%%%%%%%%%%%%%%%%%%%%%%%%%%%%
As shown in Fig.\ref{Scar-FP}(a), we identify the eigenstates bearing the scar of unstable $\pi$-mode from the maximum overlap with $\ket{\pi_+}$. The scarring of $\pi$-mode is evident from strong localization of Husimi distribution at $\phi=0$ and $\pi$, as depicted in Fig.\ref{Scar-FP}(c). Although the $\pi$-mode with energy $E=0$ coexists with the most ergodic states at the centre of the energy band, the corresponding scarred eigenstates can be distinguished from their statistical properties. According to Berry's conjecture \cite{Berry_conjecture}, the high energy eigenstates of a classically chaotic system behave as random states, although it relies on semiclassical analysis \cite{Srednicki}. For such random states, the probability distribution of their components $\eta = |\psi^i_{\nu}|^2 \mathcal{N}$, follows the well known Porter-Thomas (PT) distribution $\text{P}(\eta) = (1/ \sqrt{2\pi\eta}) \exp (-\eta/2)$ \cite{Haake}. Interestingly, the eigenstates bearing scars deviate from PT distribution, indicating the violation of Berry's conjecture \cite{sinha1,mondal}, as shown in Fig.\ref{Scar-FP}(e,f)). It is important to mention that, the magnitude of such deviation and the overlap with the corresponding coherent states, depend on the degree of scarring, which reduces with  the enhanced instability of the underlying classical dynamics. A similar analysis reveals the scarred eigenstates corresponding to symmetry unbroken unstable steady state FP-I (see Fig.\ref{Scar-FP}(b,d,f)) at energy density $E=-2$. In a similar manner, the scar of equivalent symmetry unbroken steady state FP-II can also be identified at energy density $E=+2$. 
It is to be noted that, the degree of scarring of $\pi$-mode is much less compared to that of FP-I (or FP-II), as reflected from the deviation from PT distribution, shown in Fig.\ref{Scar-FP}(e,f), since the $\pi$-mode remains always unstable and the instability is much larger compared to that of FP-I (or FP-II). Moreover, the steady state FP-I (or FP-II) is located away from the centre of the energy band comprising of maximally ergodic states.

\begin{figure}
	\centering
	\includegraphics[height=6.45cm,width=7.45cm]{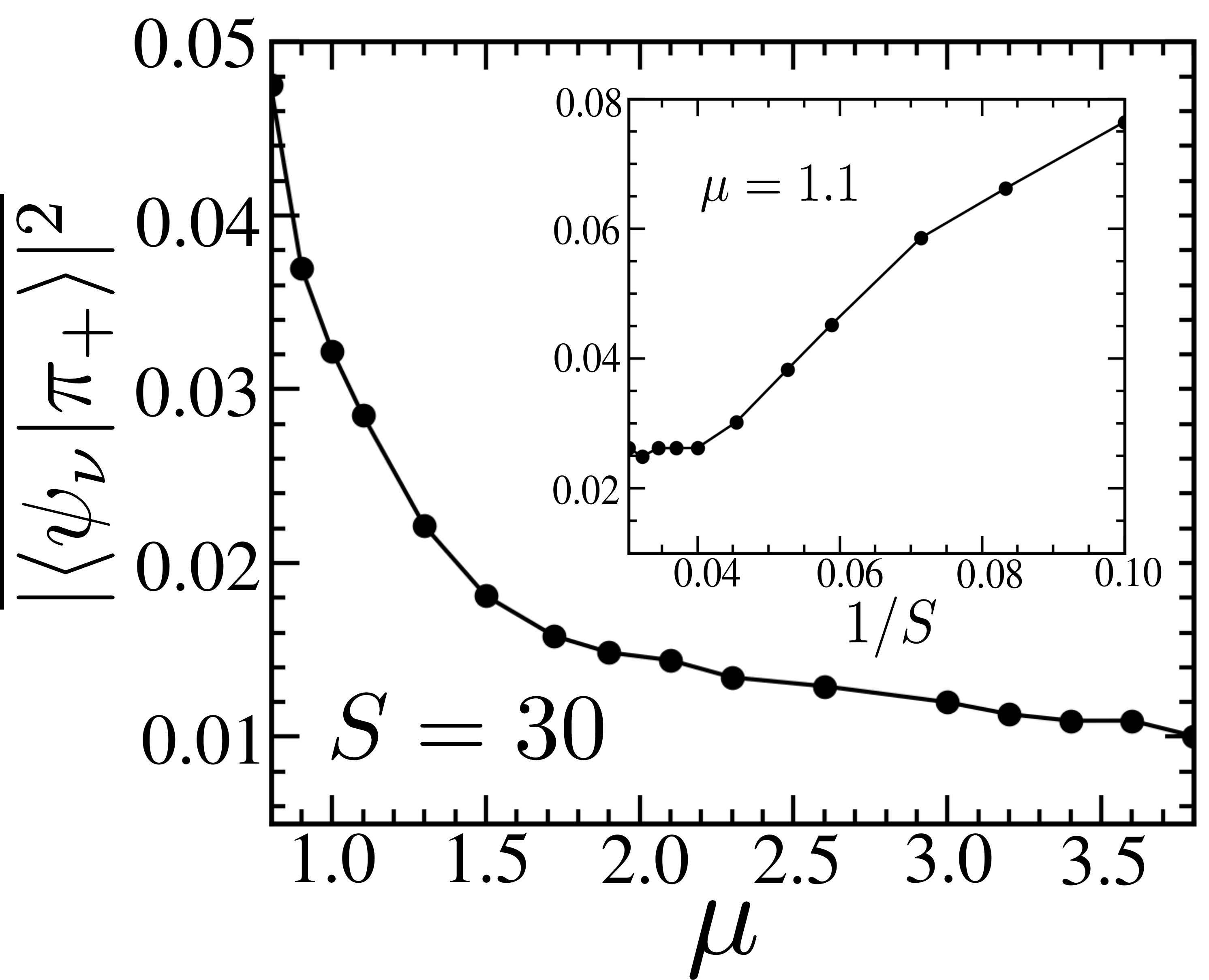}
	\caption{ Variation of average overlap $\overline{|\langle \psi_{\nu} | \pi_{+} \rangle|^2}$ of the $\ket{\pi_{+}}$ state with increasing $\mu$ for $S=30$, and with spin magnitude $S$ in the inset for $\mu = 1.1$.}
	\label{dos_increasing_S}
\end{figure}

To this end, we quantify the degree of scarring from the average overlap $\overline{|\langle \psi_{\nu}|\psi_{c} \rangle|^2}$ of the eigenstates $\ket{\psi_{\nu}}$, having significant overlap $|\langle \psi_{\nu} | \psi_{c}\rangle|^2 \gg 1/\mathcal{N}$  \cite{abanin,sinha1} ($\mathcal{N}$ being the system size) with the coherent state $\ket{\psi_{c}}$ describing the unstable FP. With increasing the coupling strength $\mu$, as the instability of the unstable FP increases, we also find that the degree of scarring decreases, as shown for scar of $\pi$-mode in Fig.\ref{dos_increasing_S}. We also study the behavior of such average overlap $\overline{|\langle \psi_{\nu}|\psi_{c} \rangle|^2}$ with increasing system size, which appears to saturate to a finite value $\gg 1/\mathcal{N}$ for sufficiently large system size, confirming the persistence of quantum scarring due to the unstable FPs, as shown for $\pi$-mode in the inset of Fig.\ref{dos_increasing_S}.
%%%%%%%%%%%%%%%%%%%%%%%%% Fig 13: Degree of scarring with S%%%%%%%%%%%%%%%%%%%%%%%%%
%%%%%%%%%%%%%%%%%%%%%%%%%%%%%%%%%%%%%%%%%%%%%%%%%%%%%%%%%%%%%%%%%%%%%%%%%%%%%%%%%

\subsection{Scars of periodic orbit and dynamical class}
\label{Scar_periodic_orbit}
\begin{figure}
	\centering
	\includegraphics[height=7.5cm,width=8.8cm]{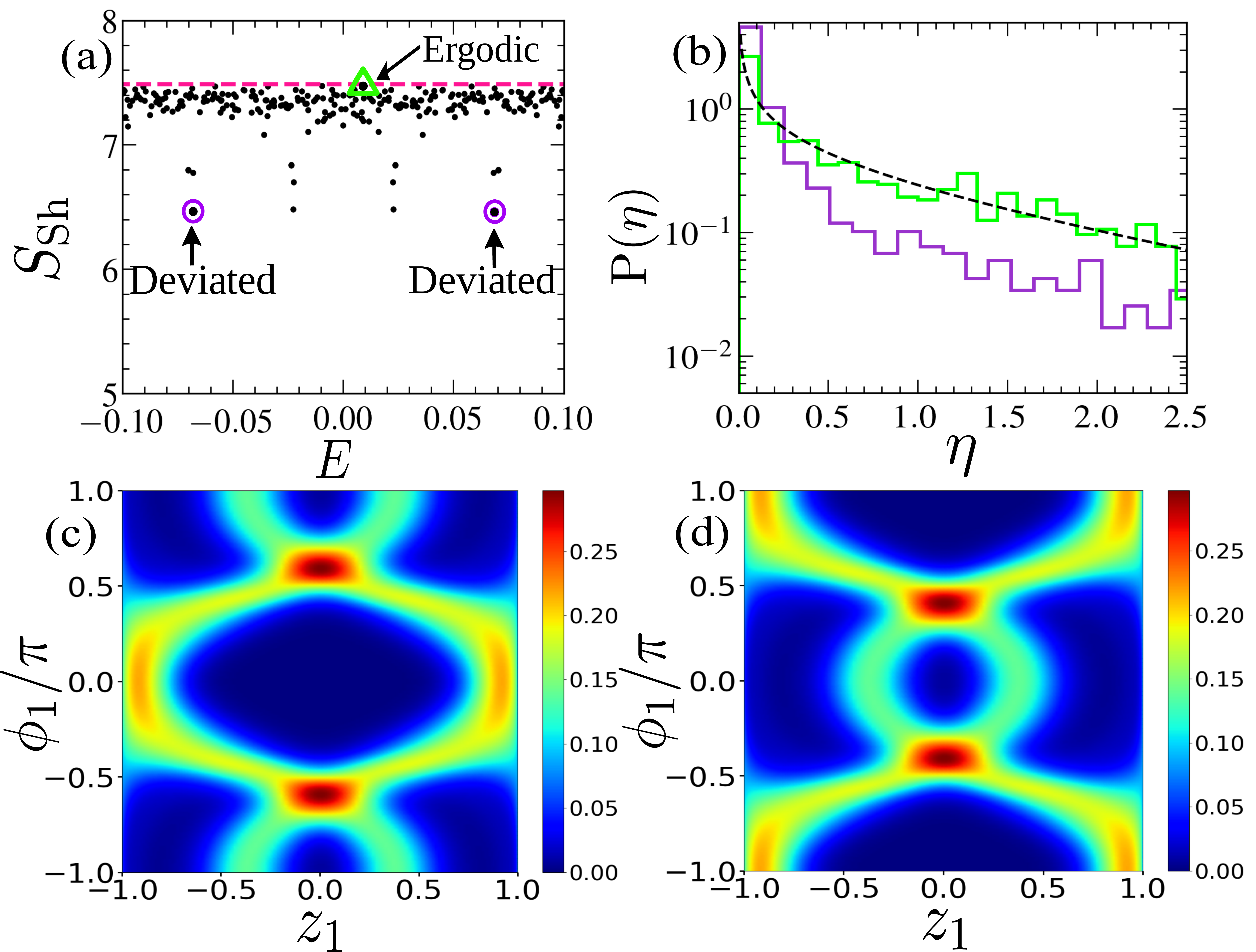}
	\caption{Identification of eigenstates with scars of unstable periodic orbits: (a) Shannon entropy $S_{\rm Sh}$ of the eigenstates around the energy density $E=0$. The pink dashed line denotes the GOE limit $\text{ln}(0.48\mathcal{N})$. (b) Comparison of distribution $P(\eta)$ for the deviated scarred states (violet line) with that of an ergodic state (green line). The black dashed line corresponds to the PT distribution. Husimi distribution for (c), (d) deviated states, marked by violet circles in (a), revealing the scar of periodic orbit. For all figures, $\mu = 1.85$.}
	\label{Scar-periodic-orbit}
\end{figure}

The second type of scars can be identified from the eigenstates maximally deviating from the ergodic limit at the centre of the energy band with $E\approx 0$, where the majority of states approach to this limit. Such behavior can be revealed from the analysis of Shannon entropy $S_{\rm Sh}=-\sum_{i}|\psi_{\nu}^{i}|^2\text{log}|\psi_{\nu}^{i}|^2$ in a small energy window of $\Delta E \sim 0.1$ around the band centre $E \approx 0$, in the ergodic regime near $\mu \approx 1.85$. As seen from Fig.\ref{Scar-periodic-orbit}(a), the Shannon entropy $S_{\rm Sh}$ for most of the states form a band like structure and approach the GOE limit ${\rm ln}(0.48\mathcal{N})$ \cite{Izrailev_1,Izrailev_2}, whereas there are a few states which have a lower value of $S_{\rm Sh}$ and deviate from it. Also, such states exhibit deviation from PT distribution (see Fig.\ref{Scar-periodic-orbit}(b)).
%although they belong to the center of the energy band with $E\approx 0$. 
The Husimi distribution corresponding to these deviated states (as indicated by circles in Fig.\ref{Scar-periodic-orbit}(a)) show a significant phase space localization, revealing the scars of periodic orbit, as depicted in Fig.\ref{Scar-periodic-orbit}(c,d). 
%%%%%%%%%%%%%%%%%%%%%%%%% Fig:14 Scar of periodic orbits%%%%%%%%%%%%%%%%%%%%%%%%%%%%
%%%%%%%%%%%%%%%%%%%%%%%%%%%%%%%%%%%%%%%%%%%%%%%%%%%%%%%%%%%%%%%%%%%%%%%%%%%%%%%%%

\begin{figure}
	\centering
	\includegraphics[height=7.5cm,width=8.6cm]{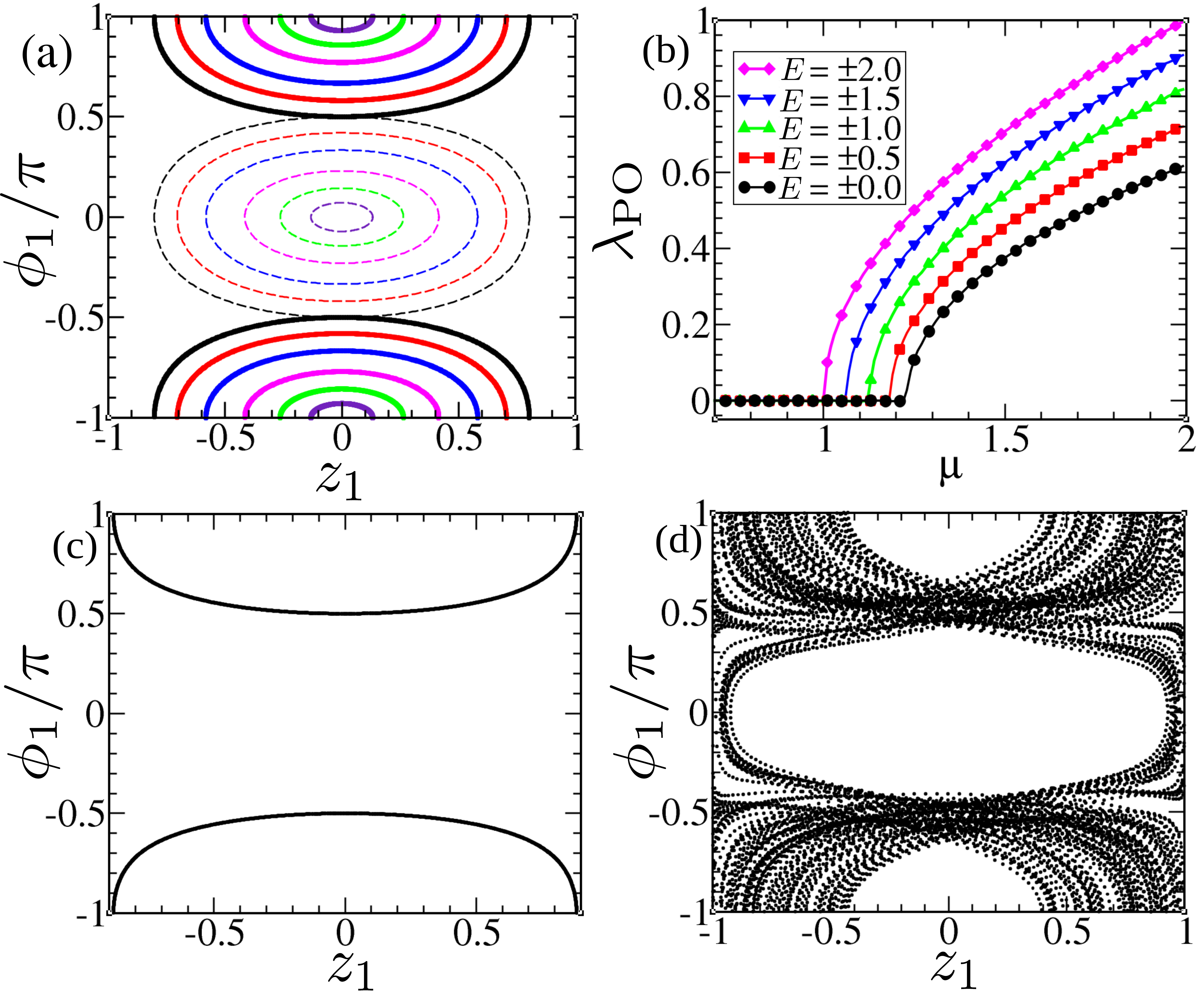}
	\caption{(a) Unstable periodic orbits corresponding to class I (dashed lines) and II (solid lines) for $\mu = 1.85$. The outermost orbits of the two classes correspond to energy $E = 0$.(b) Lyapunov exponent $\lambda_{\text{PO}}$ for the periodic orbits (with energy $E$) associated with the dynamical classes, as a function of coupling $\mu$.  Time evolution of a periodic orbit corresponding to dynamical class II with energy $E\approx 0$, in presence of a small constrain violating initial perturbation $z_-(0)=\delta z \ll 1$, (c) in the stable region with $\mu=1.2$ and (d) unstable region with $\mu=1.4$.}
	\label{Monodromy}
\end{figure}

To understand the origin of such periodic orbits, we closely analyze the classical equations of motion (EOM) given in Eq.\eqref{EOM}. As mentioned earlier, the CT model remains invariant under the exchange of two spins $(S_1 \leftrightarrow S_2)$. As a result, in terms of the redefined classical variables $z_\pm = (z_1 \pm z_2)/2$ and $\phi_\pm = (\phi_1 \pm \phi_2)/2$, the overall dynamics of the system can be categorized into two classes, namely: {\bf Class I} with $\{z_+=0;\phi_+=0\}$, and {\bf Class II} for which $\{z_-=0;\phi_-=0\}$ holds, where the dynamics is restricted on the reduced phase space of remaining variables. It can be verified from the equation of motion (EOM) in Eq.\eqref{EOM}, either of the above conditions remain valid irrespective of the coupling strength $\mu$ and the dynamics of the remaining variables reduces to that of the Lipkin-Meshkov-Glick (LMG) model \cite{LMG} with (anti)ferromagnetic interaction corresponding to class (I)II. However, the presence of initial fluctuations violating such conditions corresponding to class I and II can lead to the instability of dynamics of the corresponding classes on the reduced phase space. 
By using the EOM in Eq.\eqref{EOM} and the conditions for both the classes, the dynamics of the remaining variables can be described by,
\begin{subequations}
\begin{eqnarray}
\text{Class I:}\quad\notag\\
\dot{z}_-&=&-\sqrt{1-{z_-}^2}\sin\phi_-\\
\dot{\phi}_-&=&\frac{z_-}{\sqrt{1-{z_-}^2}}\cos\phi_-+\mu z_- 
\label{equation_z-phi-} \\
\text{Class II:}\quad\notag\\
\dot{z}_+&=&-\sqrt{1-{z_+}^2}\sin\phi_+\\
\dot{\phi}_+&=&\frac{z_+}{\sqrt{1-{z_+}^2}}\cos\phi_+-\mu z_+
\label{equation_z+phi+}
\end{eqnarray}
\end{subequations}
where the dynamical class (I)II corresponds to the LMG model with (anti)ferromagnetic interaction. From the above equations, we obtain the periodic orbits at energy $E$ in a closed form, which can be written as,
\begin{subequations}
	\begin{eqnarray}
	z_\pm(t)&=&\mathcal{A}\,\text{cn}\left(\frac{\mathcal{A}\mu}{2k}(t+t_0),k\right)\\
	\cos(\phi_\pm(t))&=&-\frac{E+\xi \mu \,{z^2_\pm(t)}}{2\sqrt{1-{z^2_\pm(t)}}} \label{jacobi-elliptic-functions}
	\end{eqnarray}
\end{subequations}
where cn is the Jacobi elliptic function with elliptic modulus $k<1$ and the constants are given by,
\begin{eqnarray}
\mathcal{A}^2&=&\,\frac{2}{\mu ^2}\left[-\xi\frac{\mu E}{2}-1+\Omega\right],\quad k^2=\frac{1}{2}\left[1-\frac{\xi E\mu/2 +1}{\Omega}\right]\nonumber,\\
t_0&=&\frac{\text{F}\left(\cos^{-1} (z_\pm(0)/\mathcal{A}),k\right)}{\Omega^{1/2}},\quad \Omega =\sqrt{\mu^2+1+\xi E\mu} \nonumber
\label{elliptic-modulus}
\end{eqnarray}
The periodic orbits corresponding to dynamical class (I)II have energy in the range $0<E<2$ ($-2<E<0$) and the parameter value $\xi=-1(+1)$, which determines the sign of the effective LMG model with (anti)ferromagnetic interaction (as seen from Eq.(\ref{equation_z-phi-}) and Eq.(\ref{equation_z+phi+})). Such orbits with different energies belonging to the dynamical classes I and  II are shown in Fig.\ref{Monodromy}(a). The periodic orbits corresponding to the classes I and II are formed around the unstable steady states FP-I and FP-II respectively, and the orbits with equal and opposite energies have identical shapes, however their centres are shifted around the corresponding FPs (see Fig.\ref{Monodromy}(a)). Interestingly, the largest orbits with energy $E=0$ belonging to the two different dynamical classes touch at $\{z_1=0,\phi_1=\pm\pi/2\}$, where the accumulation of phase space density is exhibited in the Husimi distribution of the corresponding scarred eigenstates, as shown in Fig.\ref{Scar-periodic-orbit}(c,d). The time period of such orbits with energy $E=0$ is given by,
\begin{eqnarray}
T &=& \frac{4K(k)}{(1+\mu^2)^{1/4}} \label{time-period}
\end{eqnarray}
%%%%%%%%%%%%%%%%%%%%%%%% FIG:15 instability of periodic orbits %%%%%%%%%%%%%%%%%%%%%
%%%%%%%%%%%%%%%%%%%%%%%%%%%%%%%%%%%%%%%%%%%%%%%%%%%%%%%%%%%%%%%%%%%%%%% 
where $K(k)\,=\,\text{F}(\pi/2,k)$ is the complete elliptic integral of first kind and $\text{F}(\phi,k)\,=\,\int_{0}^{\phi}dx (1-k^2\sin^2 x)^{-1/2}$ \cite{Table_integrals}.
Next, we investigate the stability of such periodic orbits by using the method of Monodromy matrix described in \cite{monodromy1,monodromy2}. Similar to the stability analysis of a phase space trajectory described in Sec.\ref{Lyapunov}, we calculate the Lyapunov exponent $\lambda_{\text{PO}}$ starting from a point on the periodic orbit over its time period $T$ and the instability of the corresponding orbit can be quantified from $\lambda_{\text{PO}}>0$. The variation of Lyapunov exponent $\lambda_{\text{PO}}$ with increasing coupling strength $\mu$ is shown in Fig.\ref{Monodromy}(b) for periodic orbits with different classical energies $E$. As seen from this figure, the largest periodic orbits with $E=0$ become unstable for $\mu>\mu_u = 1.23$, which ensures that these orbits in the ergodic regime ($\mu\approx 1.85$) are indeed unstable. However, the reminiscence of such unstable orbits are still visible in the form of scars, which is evident from the Husimi distribution of the corresponding eigenstates, as shown in Fig.\ref{Scar-periodic-orbit}(c,d). It is evident from Fig.\ref{Monodromy}(b), the smaller periodic orbits with increasing $|E|$ (from $E=0$) become successively unstable for $\mu<\mu_u$, which finally terminates at $\mu_c=1$ for $E=\pm2$ corresponding to the unstable steady states FP-I and FP-II, around which such periodic orbits are formed. Moreover, we observe that the $\lambda_{\text{PO}}$ for $E =\pm 2$ coincides with the imaginary part of the small amplitude oscillation frequency $\Im (\omega)=\sqrt{\mu-1}$, corresponding to FP-I and FP-II, as discussed previously in Sec.\ref{Classical}. From the full phase space dynamics, we identify that, the instability of such periodic orbits belonging to a particular dynamical class stems from the initial perturbations violating the constraints of the corresponding class. In Fig.\ref{Monodromy}(c,d), we elucidate such dynamics in presence of small initial perturbation deviating from a particular dynamical class, which clearly leads to instability of the periodic orbits (see Fig.\ref{Monodromy}(d)), however, the irregular trajectory remains localized around the corresponding orbits of both the dynamical classes, resembling the shape of the scar, as observed from the Husimi distributions in Fig.\ref{Scar-periodic-orbit}(c).
\subsection{Dynamical signature of scars}
\label{Dynamical_signeture_scar}
%%%%%%%%%%%%%%%%%%%%%%%% FIG:16 FOTOC of scar FP %%%%%%%%%%%%%%%%%%%%%%%%%%
\begin{figure}[H]
	\centering
	\includegraphics[height=7cm,width=8.5cm]{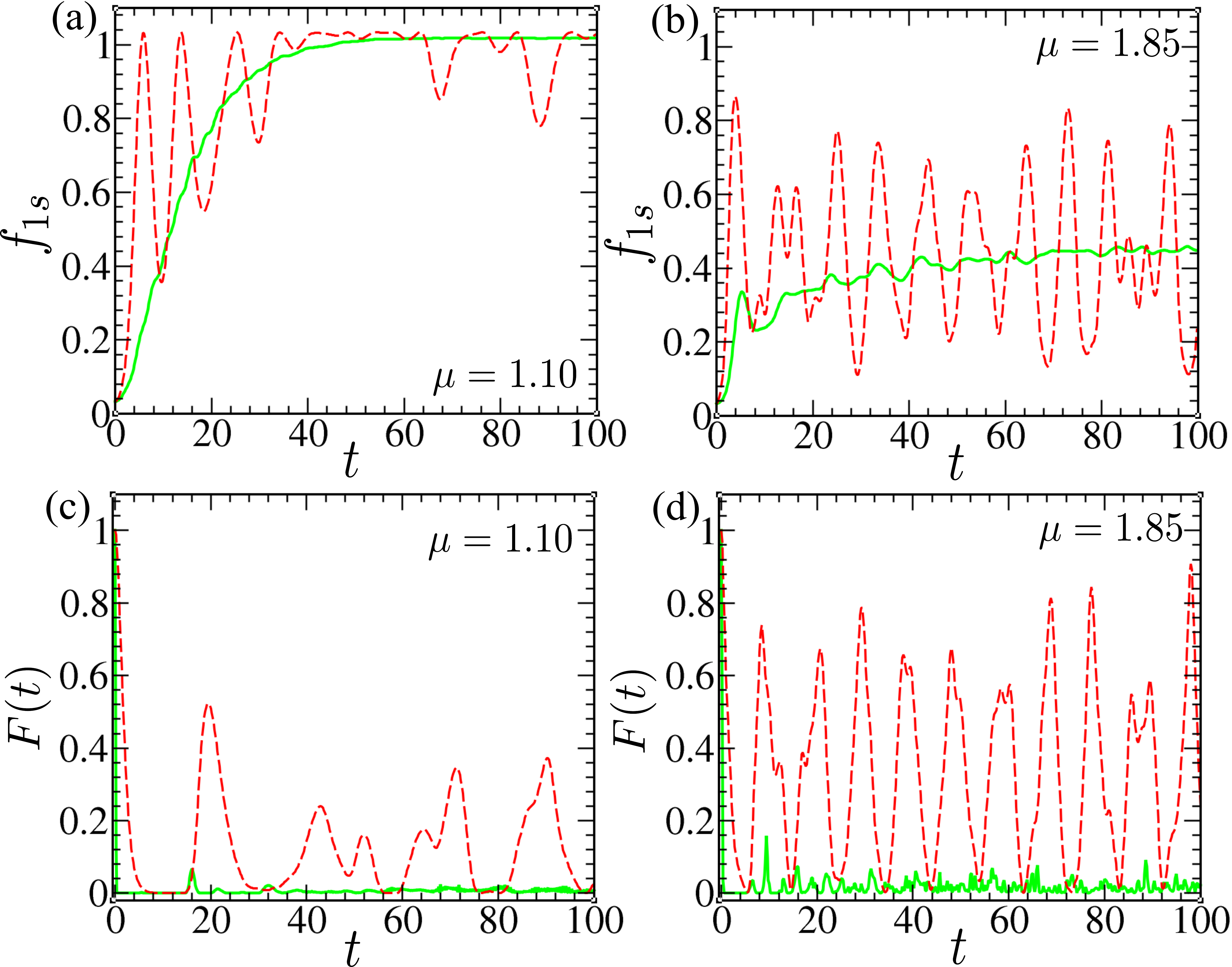}
	\caption{Dynamical signature of quantum scars: Dynamics of FOTOC $f_{1s}$ starting from the initial coherent state $\ket{\psi_c}$, representing the unstable FP of (a) $\pi-$mode and (b) FP-I, which is compared with average FOTOC $f_{1s}^E$ (green line) corresponding to same energy density $E$ of the respective unstable FPs.  Survival probability $|\langle \psi(t)|\psi_c\rangle|^2$ corresponding to unstable (c) $\pi-$mode and (d) FP-I. Green line corresponds to survival probability starting from an initial coherent state representing an arbitrary phase space point with same energy $E$ of the above mentioned FPs.}
	\label{FP-FOTOC}
\end{figure}
%%%%%%%%%%%%%%%%%%%%%%%%%%%%%%%%%%%%%%%%%%%%%%%%%%%%%%%%%%%%%%%%%%%%%%%
To this end, we discuss the detection of two types of scars from their dynamical signature using the method of out-of-time order correlators (OTOC) and survival probability. In Sec.\ref{OTOC dynamics}, we outline the OTOC technique and its extension Fidelity OTOC (FOTOC) which has been used to probe the local ergodic behavior of phase space. Since the first type of scars arise from the unstable FPs such as FP-V, I and II, it is more convenient to detect their signature from FOTOC dynamics. Given that, the FOTOC is directly related to the fluctuation of the corresponding operator, here we investigate the behavior of total spin fluctuation $f_{1s}$, defined in Eq.\eqref{FOTOC} to probe the presence of first type of scar. 
Starting from an initial coherent state representing the unstable FPs, the dynamics of the total spin fluctuations $f_{1s}$ exhibits oscillations with large magnitude, as depicted in Fig.\ref{FP-FOTOC}(a,b). For comparison, we also compute the mean value of FOTOC $f^{E}_{1s}$ (discussed in Sec.\ref{OTOC dynamics}) averaged over an ensemble of initial coherent states with energy density $E$ corresponding to that of the unstable FP. As observed from Fig.\ref{FP-FOTOC}(a,b), the $f^{E}_{1s}$ exhibits a smooth behavior and saturation in absence of oscillations, in contrast to the oscillatory behavior of $f_{1s}$ for unstable FPs. We also observed that the oscillations in the dynamics of $f_{1s}$ decrease as the overlap of the initial coherent state with that corresponding to the unstable FP reduces. Since the quantum scars lead to the deviation from ergodicity and retain the memory of the initial states, they can also be detected from the survival probability $F(t)=|\langle \psi(t)|\psi(0)\rangle|^2$, describing the overlap of time evolved state $\ket{\psi(t)}$ with the initial state $\ket{\psi(0)}$. 
The time evolution of survival probability starting from the initial coherent state representing the FP-V and FP-I are shown in Fig.\ref{FP-FOTOC}(c,d). The revival phenomena as observed in the survival probability, for the initial coherent states corresponding to the unstable steady states, is a characteristic of such scarred states \cite{Turner}, which is absent in the ergodic evolution. Also, the survival probability at long time differs from the GOE limit, indicating the deviation from ergodicity.
%%%%%%%%%%%%%%%%%%%%%%%% FIG:17 OTOC of second scar %%%%%%%%%%%%%%%%%%%%%%
\begin{figure} [H]
	\centering
	\includegraphics[height=5.5cm,width=7cm]{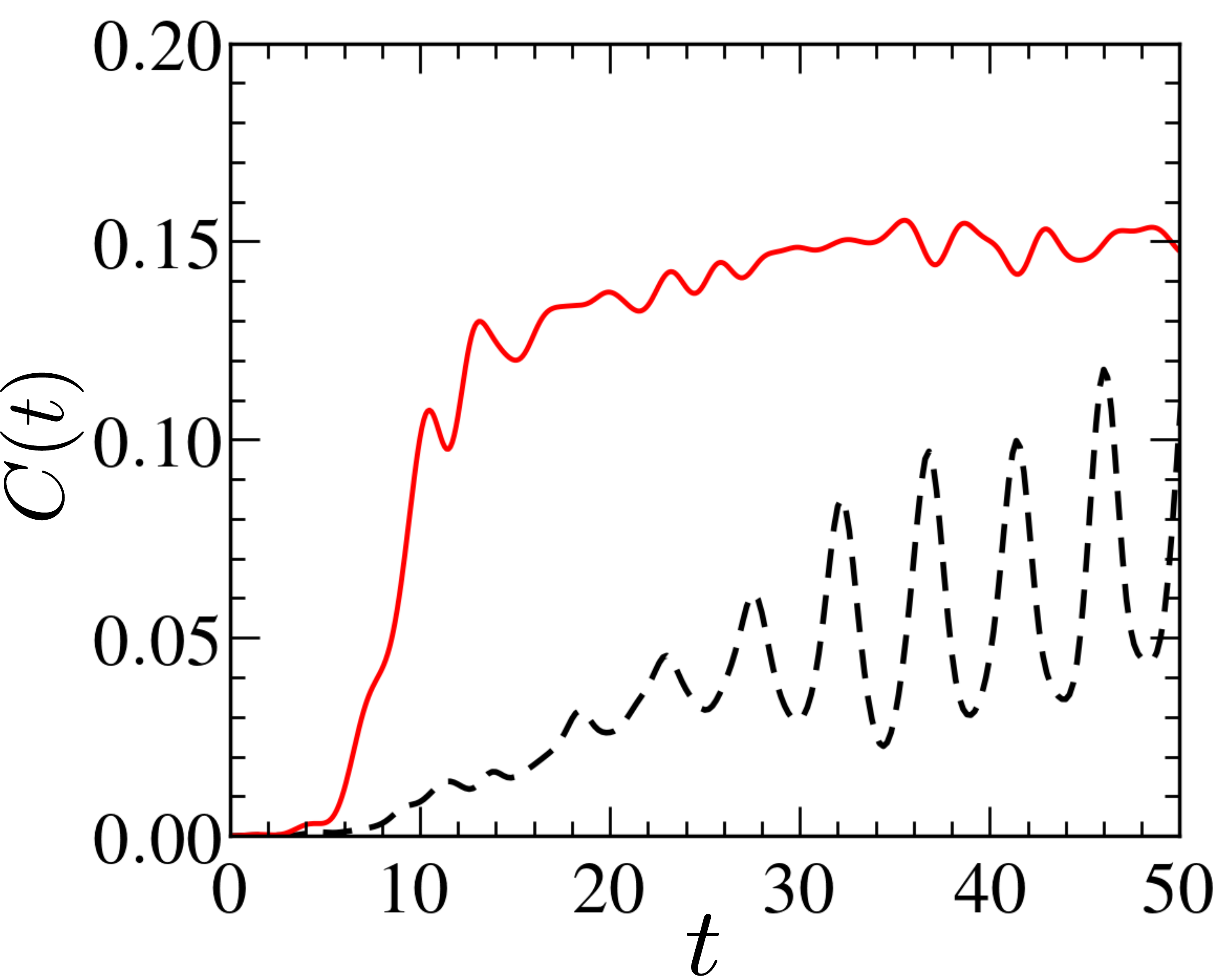}
	\caption{Signature of scar of unstable periodic orbit obtained from OTOC dynamics. Time evolution of $C(t)$ corresponding to two different initial density matrix, $\hat{\rho}_{0}$ (black dashed line) and $\hat{\rho}_{mc}$ (red solid line), as described in the main text, for $\mu = 1.85$. Dynamics of OTOC corresponding to scarred states (black dashed line) exhibits oscillation with periodicity close to time period $T$ (see Eq.\eqref{time-period}) of the unstable periodic orbit.}
	\label{periodic_orbit_FOTOC}
\end{figure}
%%%%%%%%%%%%%%%%%%%%%%%%%%%%%%%%%%%%%%%%%%%%%%%%%%%%%%%%%%%%%%%%%%%%%%%
We detect the dynamical signature of second type of scars corresponding to the unstable periodic orbits using the unequal time commutator of spins, $C(t) = -{\rm Tr}\hat{\rho}_0[\hat{S}_z(t)/S,\hat{S}_z(0)/S]^2$, which is also discussed in Sec.\ref{OTOC dynamics}. Next, we study the time evolution of $C(t)$ for initial density matrix $\hat{\rho}_0=\ket{\psi(0)}\bra{\psi(0)}$ constructed from the eigenstates carrying such scars, which are located near the band centre, where the majority of states are ergodic (see Fig.\ref{Scar-periodic-orbit}(a)). For comparison, we also investigate $C(t)$ evaluated for initial microcanonical density matrix $\hat{\rho}_{mc}=\sum_{j}
\ket{\psi_j}\bra{\psi_j}/\mathcal{N}_{\Delta}$ constructed from $\mathcal{N}_{\Delta}$ ergodic states within a small range of energy density $\Delta E$ at the centre of the band. As seen from Fig.\ref{periodic_orbit_FOTOC}, the unequal time commutator $C(t)$ corresponding to the scarred eigenstate exhibits oscillatory behavior, in contrast to the $C(t)$ showing a rapid growth and saturation in absence of  oscillations for the ergodic states.  The oscillatory behavior observed in FOTOC, as well in the unequal time commutator can be considered as a detectable signature for both types of scars.
\section{conclusion}
\label{Conclusion} 
To summarize, we have investigated various aspects of the coupled top (CT) model, which includes analysis of different types of quantum transitions and ergodic behavior of the system as well its deviation, revealing their connection with the underlying collective spin dynamics.

Ferromagnetic interaction between the spins lead to the quantum phase transition (QPT) at a critical coupling strength $\mu_c =1$, which gives rise to ferromagnetic ordering of the ground state. Moreover, at the same coupling, the highest excited state undergoes a dynamical transition with anti-ferromagnetic ordering. For both the transitions, the lowest collective frequency vanishes at the critical point $\mu_c$ as $\sqrt{|\mu-\mu_c|}$. Such gapless excitation at the critical point can lead to a large spin fluctuation of the ground state, which has been analyzed using Holstein-Primakoff approximation.
Above $\mu_c$, both QPT and dynamical transition are accompanied by two `excited state quantum phase transitions' (ESQPT) characterized by the singularities in density of states at critical energy densities. Such energy densities associated with ESQPT correspond to the symmetry unbroken unstable steady states, which separate the symmetry unbroken eigenstates at the  centre of the energy band from the symmetry broken states at the band edge. We also derive an effective potential which provides a pictorial description of both QPT and ESQPT.

Although, the CT model is integrable in both the extreme limits of coupling, $\mu \ll 1$ and $\mu \gg 1$, the classical analysis reveals a transition to chaos in an intermediate range of coupling strength above $\mu_c$. At the quantum level, the overall chaotic behavior is detected from the spectral statistics, however a closer investigation  reveals the existence of non-ergodic multifractal eigenstates across the energy band. We quantify the degree of ergodicity of eigenstates from relative entanglement entropy as well as multifractal dimensions. Both the quantities exhibit similar behavior across the energy band, revealing variation of degree of ergodicity with energy density. In the maximally chaotic regime, the eigenstates near the band centre approach ergodic limit, whereas the degree of ergodicity of states decreases toward the band edge. Such ergodic behavior across the energy band can be probed from non-equilibrium dynamics by a suitably chosen initial state at a given energy density, 
which is also supplemented by corresponding phase space dynamics, elucidating the connection between phase space mixing and ergodicity of its quantum counterpart.

Finally, we discuss another source of deviation from ergodicity due to the formation of two types of quantum scars, which arises as a reminiscence of unstable fixed points and periodic orbits. We present systematic methods to identify such eigenstates bearing quantum scars, which are found to be violating Berry's conjecture in contrast to the ergodic states. 
More importantly, we have shown how the energy dependent degree of ergodicity can be probed and dynamical signature of quantum scars can be detected by using a newly developed technique known as `out-of-time-order correlator', which has already been implemented in cold ion experiments \cite{NMR,Trapped_ion}.  

In conclusion, the present work provides a detailed discussion of a rich variety of phenomena observed in a simple collective spin model known as coupled top model, which also elucidates dynamical route to ergodicity, and formation of quantum scars in presence of interaction, as well its detection.

%\section{Acknowledgments}

%%%%%%%%%%%%%%%%%%%%%%%%%%%%% APPENDIX %%%%%%%%%%%%%%%%%%%%%%%%%%%%%%%%%%%%%%%%%
\appendix 

%%%%%%%%%%%%%%%%%%%%%%%% HOLSTEIN PRIMAKOFF %%%%%%%%%%%%%%%%%%%%%%%%%%%%%%%%
\section{Holstein Primakoff transformation}
\label{Holstein_primakoff}
For large magnitude of spins $S\gg 1$, the coupled top (CT) model can be analyzed semiclassically by using Holstein Primakoff (HP) approximation \cite{HPT}, where $1/S$ is considered to be a small parameter. When the classical spin vector is aligned along the $z$-axis, the corresponding spin operators in HP transformations can be written as,
\begin{subequations}
	\begin{eqnarray}
	\hat{\widetilde{S}}_{iz}&=& S-\hat{a}^{\dagger}_{i}\,\hat{a}_{i}\\ \hat{\widetilde{S}}_{i+}&=&\sqrt{2S-\hat{a}^{\dagger}_{i}\,\hat{a}_{i}}\,\,\hat{a}_{i}\\ \hat{\widetilde{S}}_{i-}&=&\hat{a}^{\dagger}_{i}\sqrt{2S-\hat{a}^{\dagger}_{i}\,\hat{a}_{i}}
	\end{eqnarray}
\end{subequations}
where the index $i=1,2$ represents two spins of CT model and $\hat{a}_{i}$ is bosonic annihilation  operator describing the quantum fluctuations around the classical spin. As discussed in Sec.\ref{Classical}, corresponding to the classical spin configuration of the ground state, both the spins of CT model lie in the $x-z$ plane making an angle $\theta_i$ with the $z$-axis which depends on the coupling $\mu$. For such orientation of the classical spin vectors, corresponding spin operators in HP representation can be obtained by performing a rotation about the $y$-axis,
\begin{subequations}
	\begin{eqnarray}
	\hat{S}_{ix}&=&  \hat{\widetilde{S}}_{ix}\cos\theta_i+ \hat{\widetilde{S}}_{iz}\sin\theta_i\\
	\hat{S}_{iz}&=&  \hat{\widetilde{S}}_{iz}\cos\theta_i- \hat{\widetilde{S}}_{ix}\sin\theta_i
	\end{eqnarray}
\end{subequations}
Using these spin operators, the Hamiltonian given in Eq.\eqref{Coupled Top} can be written in a series of $1/S$,
\begin{eqnarray}
\hat{\mathcal{H}}= S\mathcal{H}_0+ \sqrt{S}\hat{\mathcal{H}}_1+\hat{\mathcal{H}}_2+O(1/\sqrt{S})
\end{eqnarray}
where the dominant term $\mathcal{H}_0$ represents the classical energy density. Minimizing the classical energy $\mathcal{H}_0$, we obtain the following  conditions,
\begin{eqnarray}
\cos\theta_i &=&
\begin{dcases*} 
0 & for  $\mu < 1$ \\ 
\pm \sqrt{1-1/\mu^2} & for $\mu \ge 1$
\end{dcases*} \label{HPT}
\end{eqnarray}
which represents the spin orientation of the ground state. Next term in the Hamiltonian $\hat{\mathcal{H}}_1$ vanishes for such ground state spin configuration. Finally, the Hamiltonian $\hat{\mathcal{H}}_2$ describing the quantum fluctuations can be written as,
\begin{eqnarray}
\hat{\mathcal{H}}_2=\epsilon\,\hat{a}^{\dagger}_{1}\,\hat{a}_{1}+\epsilon \,\hat{a}^{\dagger}_{2}\,\hat{a}_{2}-\frac{\Gamma}{2}(\hat{a}_1+\hat{a}_1^{\dagger})(\hat{a}_2+\hat{a}_2^{\dagger})
\label{H before diagonalization}
\end{eqnarray}
where the parameters are given by,
\begin{eqnarray}
%\begin{dcases*} 
\epsilon &=& 1 \,\, , \,\,\Gamma = \mu \qquad \text{for}  \,\,\mu < 1 \nonumber\\ 
\epsilon &=& \mu \,\,, \,\,\Gamma = 1/\mu \quad \text{for} \,\,\mu \ge 1 \nonumber
%  \end{dcases*} \label{HPT}
\end{eqnarray}
The excitation energies above the ground state can be obtained by diagonalizing the Hamiltonian $\hat{\mathcal{H}}_2$, which can be achieved by canonical transformation of bosonic operators \cite{Emary_Brandes},
\begin{subequations}
	\begin{eqnarray}
	\hat{b}_1^{\dagger}=\frac{(\omega_-+\epsilon)(\hat{a}_1^{\dagger}-\hat{a}_2^{\dagger})+(\omega_--\epsilon)(\hat{a}_1-\hat{a}_2)}{2\sqrt{2\,\epsilon\,\omega_-}}\\
	\hat{b}_2^{\dagger}=\frac{(\omega_++\epsilon)(\hat{a}_1^{\dagger}+\hat{a}_2^{\dagger})+(\omega_+-\epsilon)(\hat{a}_1+\hat{a}_2)}{2\sqrt{2\,\epsilon\,\omega_+}}
	\end{eqnarray}
	\label{Bogoliubov transformation}
\end{subequations}
where the new set of operators satisfy the commutation relation $[\hat{b}_i,\hat{b}^{\dagger}_j]=\delta_{ij}$. The diagonal form of the Hamiltonian is given by,
\begin{eqnarray}
\hat{\mathcal{H}}_2&=& \sqrt{\epsilon^2 - \Gamma \epsilon}\,\,\hat{b}^{\dagger}_1\,\hat{b}_1+\sqrt{\epsilon^2 + \Gamma \epsilon}\,\,\hat{b}^{\dagger}_2\,\hat{b}_2 + E_0 \nonumber \\
&=& \omega_-\,\hat{b}^{\dagger}_1\,\hat{b}_1+\omega_+\,\hat{b}^{\dagger}_2\,\hat{b}_2 + E_0
\end{eqnarray}
where $\omega_{\pm}$ are the frequency of the collective modes, which matches exactly with the small amplitude oscillation frequencies (Eq.\eqref{Excitation frequency}) obtained from linear stability analysis. The zero point energy due to quantum fluctuation is given by $E_0=\left[\sqrt{\epsilon^2+\Gamma\epsilon}+\sqrt{\epsilon^2-\Gamma\epsilon}-2\epsilon\right]/2$. The correction due to finite size effect can be obtained by considering $1/S$ terms systematically \cite{Vidal_PRA}. 

\begin{figure} [H]
	\centering
	\includegraphics[height=5cm,width=6.5cm]{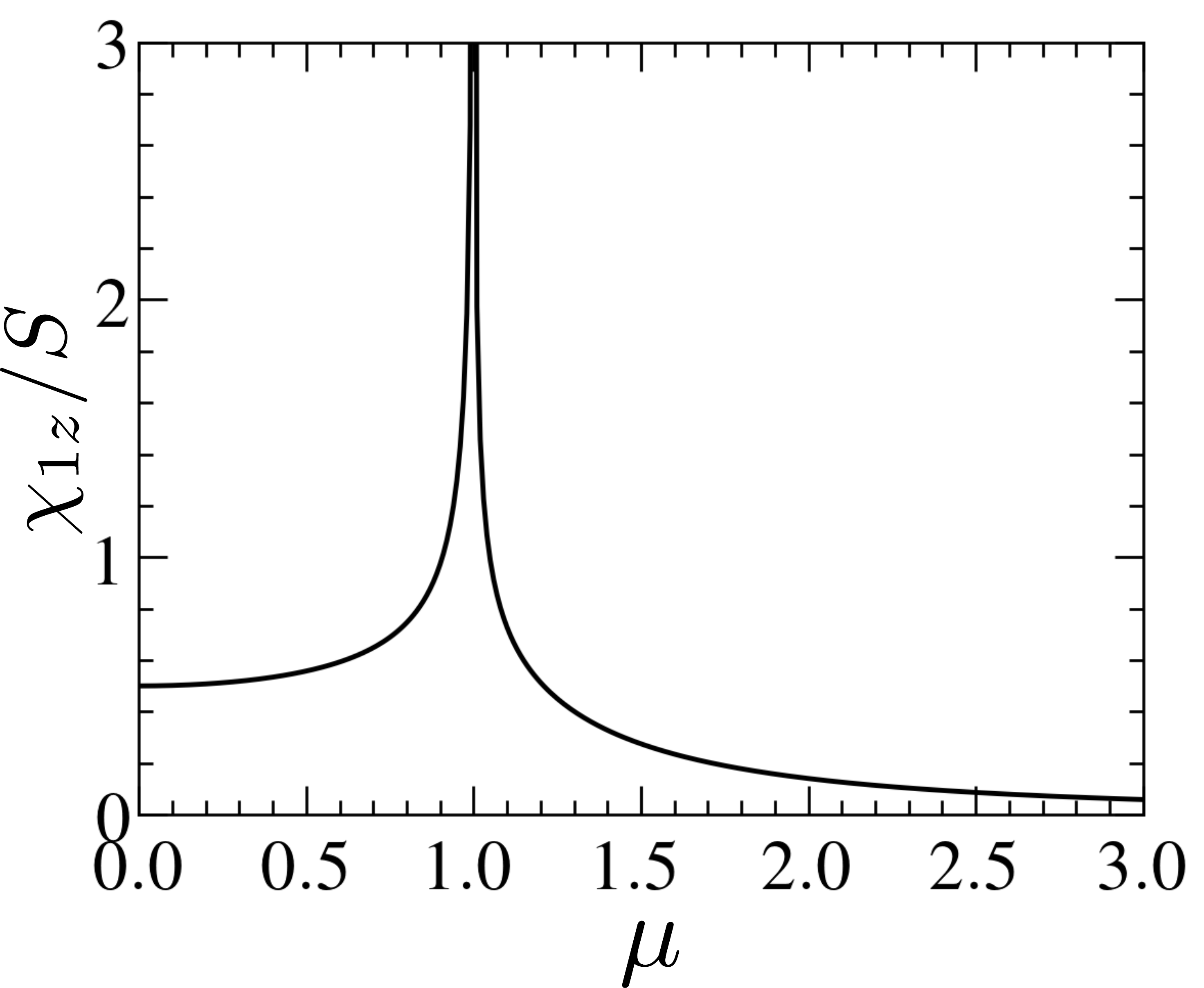}
	\caption{Signature of QPT from the singular behavior of spin fluctuation $\chi_{1z}/S$ at the critical coupling strength $\mu_c$.}
	\label{Spin-Fluctuation}
\end{figure}

Vanishing of the energy gap leads to the enhanced quantum fluctuations at the critical point, which can be observed in the spin fluctuation.
Within the HP transformation, we obtain the spin fluctuation of the ground state $\chi_{1z} = \langle \hat{S}_{1z}^2\rangle - \langle \hat{S}_{1z}\rangle ^2$ corresponding to a particular spin sector ($S_1$), 
\begin{eqnarray}
\chi_{1z}&=&\frac{S}{2}\cos^2\theta_1\left[\frac{(\epsilon- \omega_{-})^2}{\epsilon\,\omega_-}+\frac{(\epsilon-\omega_+)^2}{\epsilon\,\omega_+}\right]\nonumber\\
&&+\frac{S}{4}\epsilon\,\,\sin^2\theta_1 \,\,\left[\frac{1}{\omega_-}+\frac{1}{\omega_+}\right]
\label{Fluctuation_Sz}
\end{eqnarray} 
which diverges at the critical coupling strength $\mu_c$, as shown in Fig.\ref{Spin-Fluctuation}.
%%%%%%%%%%%%%%%%%% Fig:A1 Spin fluctuation for ground state%%%%%%%%%%%
%%%%%%%%%%%%%%%%%%%%%%%%%%%%%%%%%%%%%%%%%%%%%%%%%%%%%%%%%%%%%%%%%%%%%%%%
%%%%%%%%%%%%%%%%%%%%%%%%Effective Potential%%%%%%%%%%%%%%%%%%%%%%%%%%%%%
\section{Effective potential of the Coupled top model}
\label{effective_potential}
\begin{figure}
	\centering
	\includegraphics[height=7.7cm,width=8.4cm]{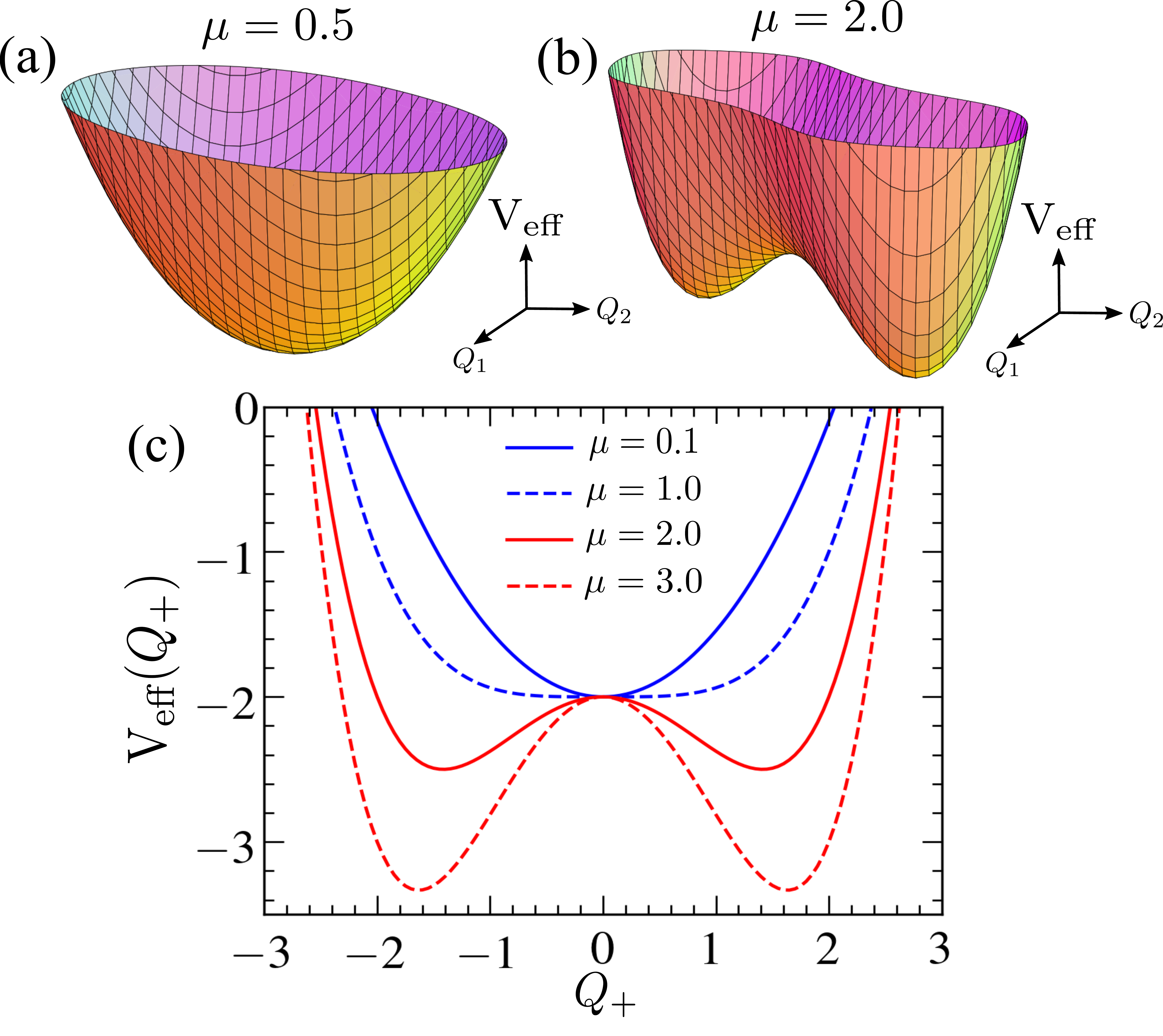}
	\caption{  Effective potential (a) before and (b) after QPT, exhibiting the symmetry breaking phenomena. (c) Cross section of effective potential with $Q_{-}=0$ for different values of $\mu$. Blue (red) lines denote the potential for $\mu\le\mu_c$ ($\mu>\mu_c$).}
	\label{Effective potential}
\end{figure}

Similar to the Landau free energy for continuous phase transition, the coupled top model can equivalently be described by an effective potential, which elucidates the quantum phase transition, and its connection with ESQPT.
Following the prescription of HP transformation, we define the canonically conjugate variables $Q_i,P_i$ in terms of the bosonic creation $\hat{a}_i^{\dagger}$ and annihilation $\hat{a}_i$ operators as,
\begin{eqnarray}
Q_i = \frac{\hat{a}_i^{\dagger}+\hat{a}_i}{\sqrt{S}},\hspace{0.5cm}P_i =i\,\frac{\hat{a}_i^{\dagger}-\hat{a}_i}{\sqrt{S}}
\end{eqnarray}
which can be treated classically in the limit $S\rightarrow \infty$. Using the HP transformation, the spin components can be written in terms of these variables as,
\begin{subequations}
	\begin{eqnarray}
	\tilde{S}_{ix}&=& \frac{S}{\sqrt{2}}Q_i\sqrt{1-\frac{Q_i^2+P_i^2}{8}}\\
	\tilde{S}_{iy}&=& \frac{S}{\sqrt{2}}P_i\sqrt{1-\frac{Q_i^2+P_i^2}{8}}\\
	\tilde{S}_{iz}&=& S\left(1-\frac{Q_i^2+P_i^2}{4}\right)
	\end{eqnarray} 
	\label{Spin_QP}
\end{subequations}
In the symmetry unbroken phase with $\mu<\mu_c$, spins are aligned along the $x-$axis in the ground state of the system. For such classical spin orientation we perform a spin rotation around the y-axis, which yields $\tilde{S}_{ix} = S_{iz}$ and $\tilde{S}_{iz} = -S_{ix}$ and in terms of the classical variables $Q_i,P_i$ the corresponding classical Hamiltonian (scaled by $S$) can be written as,
\begin{small}
	\begin{eqnarray}
	\mathcal{H}_{cl}&=&-2+\frac{Q_1^2+P_1^2}{4}+\frac{Q_2^2+P_2^2}{4}\nonumber\\
	&&-\frac{\mu}{2}Q_1Q_2\sqrt{1-\frac{Q_1^2+P_1^2}{8}}\sqrt{1-\frac{Q_2^2+P_2^2}{8}}.
	\label{H_QP}
	\end{eqnarray}
\end{small}
Minimization of the energy leads to the conditions $P_1=P_2=0$ and thus we obtain the effective potential for CT model,
\begin{small}
	\begin{eqnarray}
	\text{V}_{\text{eff}}=-2+\frac{Q_1^2+Q_2^2}{4}-\frac{\mu}{2}Q_1Q_2\sqrt{1-\frac{Q_1^2}{8}}\sqrt{1-\frac{Q_2^2}{8}}.
	\label{V_eff}
	\end{eqnarray}
\end{small}
It can be seen from Fig.\ref{Effective potential}(a), below $\mu_c$, the effective potential $\text{V}_{\text{eff}}$ has a single minimum at $Q_1=Q_2=0$, which corresponds to the symmetry unbroken steady state FP-I with energy $E=-2$. Whereas above the phase transition $\mu>\mu_c$, the potential changes its shape to double well structure (see Fig.\ref{Effective potential}(b)), which clearly captures the symmetry breaking phenomenon associated with QPT. The energy of the two minima of this potential matches with that of the symmetry broken steady state FP-III. 
Above the QPT with $\mu>\mu_c$, the point $Q_1=Q_2=0$ corresponds to the maximum of the potential barrier in $\text{V}_{\text{eff}}$, which schematically describes the ESQPT with critical energy density $E_c=-2$, separating the symmetry broken states (within the double well) from the symmetry unbroken sector above the barrier. 
%%%%%%%%%%%%%%%%%%%%%%%% Fig:B1 Effective potential %%%%%%%%%%%%%%%%%%%%%%%%%%%
%%%%%%%%%%%%%%%%%%%%%%%%%%%%%%%%%%%%%%%%%%%%%%%%%%%%%%%%%%%%%%%%%%%%%%%%%
The Hamiltonian given in Eq.\eqref{H_QP} remains invariant under the exchange of the variables ($Q_1\leftrightarrow Q_2,P_1\leftrightarrow P_2$). Therefore, we redefine the variables as, $Q_{\pm}=(Q_1\pm Q_2)/2$ and the effective potential (Eq.\eqref{V_eff}) can be rewritten in terms of these new variables. It can be shown, the minimization of this redefined potential leads to the condition $Q_-=0$, which is in accordance with the fact that, due to the ferromagnetic ordering of the ground state, the spins are aligned. Using the condition $Q_-=0$, the potential can be written in terms of $Q_+$ as,
\begin{eqnarray}
\text{V}_{\text{eff}}(Q_+)=-2+\frac{Q_+^2}{2}-\frac{\mu}{2}\left(Q_+^2-\frac{Q_+^4}{8}\right)
\end{eqnarray}
The nature of this potential as a function of $Q_+$, in the different coupling regime is presented in the Fig.\ref{Effective potential}(c), which capture the signature of quantum phase transition. Furthermore, It is evident from the Fig.\ref{Effective potential}(c) that, the ESQPT corresponds to the barrier between the two well, and the barrier height corresponds to the critical energy density $E_c=-2$.
%%%%%%%%%%%%%%%% SIGNATURE OF PHASE TRANSITION FROM FRACTAL DIMENSION %%%%%%%%%%%%
\section{Signature of Quantum phase transition in the multifractal dimension}
\label{multifractal_dimension_QPT}
\begin{figure} [H]
	\centering
	\includegraphics[height=5.5cm,width=6.5cm]{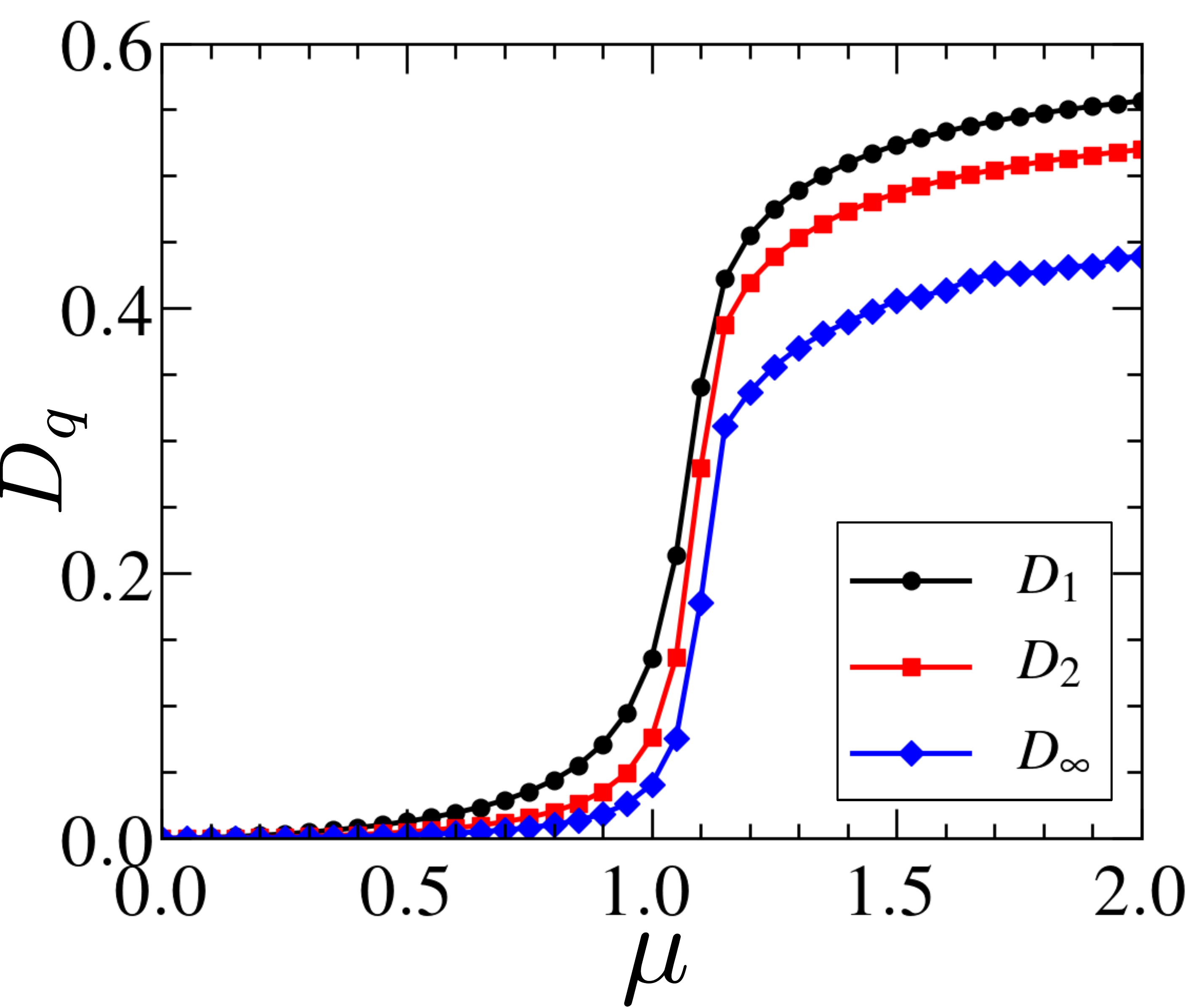}
	\caption{Signature of QPT from  the multifractal dimension $D_q$ of the ground state as a function of $\mu$.}
	\label{D1-D2}
\end{figure}

The change in the structure of ground state wavefunction across the quantum critical point can be captured from the sudden change in fractal dimension $D_q$, which signifies the occurrence of quantum phase transition. Before the transition ($\mu<\mu_c$), both spins of the CT model are aligned along the $x$-axis, as a result, a few eigenstates of $\hat{S}_{ix}$ operators with large eigenvalues participate in the ground state. Hence, the ground state remains localized in the basis $\ket{m_{1x},m_{2x}}$, where $m_{ix}$ is the eigenvalue of $\hat{S}_{ix}$. Consequently, before QPT, the multifractal dimension $D_q$ computed in the above basis yields small value indicating localization. On the other hand, the spin orientation changes above the critical coupling $\mu_c$ and more number of eigenstates contribute to the ground state.
Such change in the structure of the ground state is reflected from rapid growth in multifractal dimension $D_q$ across the QPT as shown in Fig.\ref{D1-D2}. 
%%%%%%%%%%%%%%%%%% Fig: C1 Dq vs mu for ground state%%%%%%%%%%%
%%%%%%%%%%%%%%%%%%%%%%%%%%%%%%%%%%%%%%%%%%%%%%%%%%%%%%%
%%%%%%%%%%%%%%%%%%%%Integrability at large coupling strength%%%%%%%%%%%%%%%%%%%%%
\section{Integrability at large coupling strength}
\label{integrability_large_coupling}
\begin{figure}[H]
	\centering
	\includegraphics[height=3.85cm,width=8.5cm]{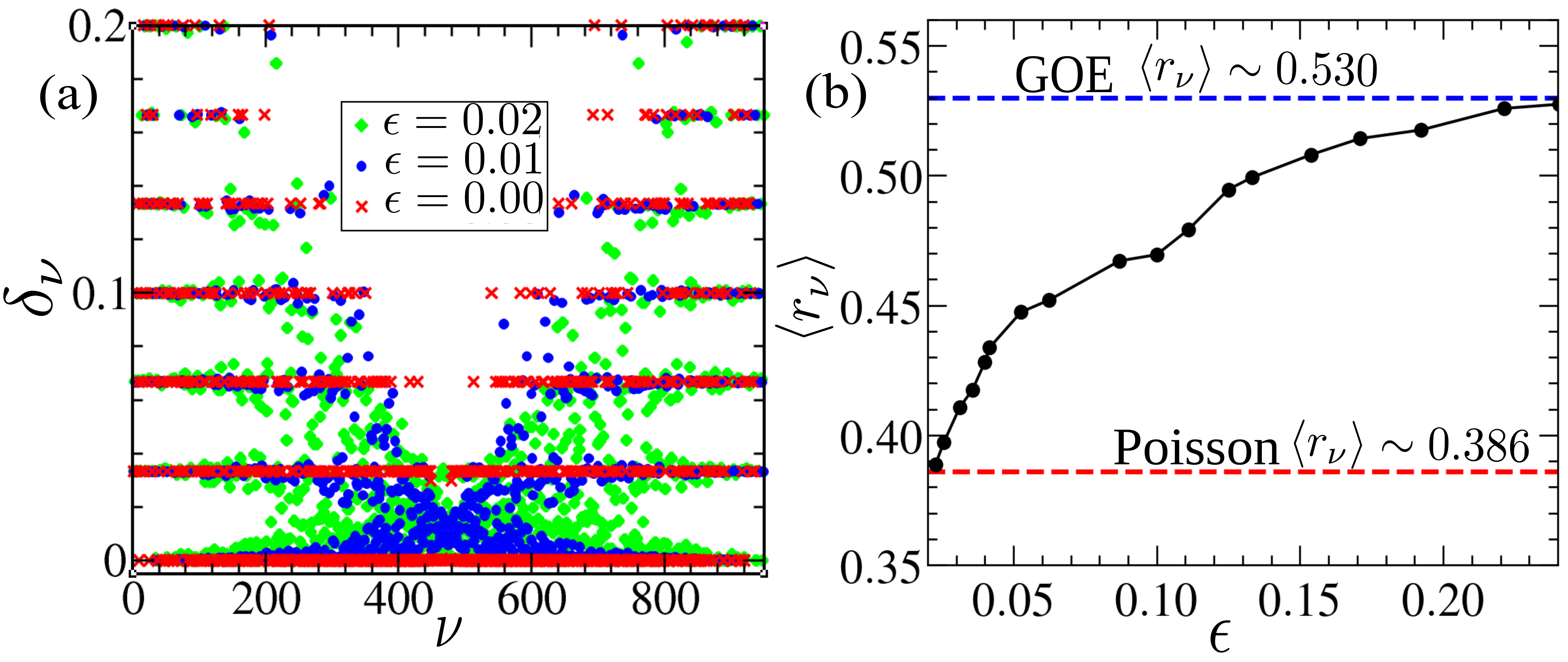}
	\caption{(a) Level spacing $\delta_{\nu}$ with increasing index $\nu$ of ordered eigenvalues (of even symmetry sector), for different values of $\epsilon$. The eigenstates near both edges of the energy band exhibit regular spacing $\delta_\nu$, as the integrable limit is approached with decreasing $\epsilon$. (b) Variation of average ratio of consecutive level spacing  $\langle r_{\nu}\rangle$ with $\epsilon$, exhibiting crossover from GOE to Poisson statistics. Red (blue) dashed lines denote the Poisson (GOE) limit of $\langle r_{\nu} \rangle$.}
	\label{label spacing large mu}
\end{figure}

The coupled top model becomes integrable at two extreme limits of coupling strength $\mu$. In the presence of weak interaction ($\mu \rightarrow 0$), two spins precess independently around the $x-$axis and the system becomes integrable. At the quantum level, this is manifested by Poissonian level spacing distribution in the small coupling regime $\mu<\mu_c$. Moreover, in the opposite limit ($\mu\gg 1$) where the interaction between the two spins becomes large compared to the precession term, the model is nearly integrable. To study the system in this extreme limit, we scale the Hamiltonian (given in Eq.\eqref{Coupled Top}) by $\mu$, which can be written as,
\begin{eqnarray}
\hat{\mathcal{H}} = -\epsilon(\hat{S}_{1x}+\hat{S}_{2x})-\frac{1}{S}\hat{S}_{1z}\hat{S}_{2z} \label{integrable_limit}
\end{eqnarray}
where $\epsilon=1/\mu$ and $\epsilon \rightarrow 0$ for $\mu\rightarrow \infty$.
The above Hamiltonian can be written classically in terms of the collective coordinates as,
\begin{eqnarray}
\mathcal{H}_{cl} = -\epsilon\left[\sqrt{1-z_1^2}\cos{\phi_1}+\sqrt{1-z_2^2}\cos{\phi_2}\right]-z_1z_2\hspace{0.3cm}
\label{semiclassical_ham_integrable}
\end{eqnarray}
%%%%%%%%%%%%%%%%%%%%%%%%%%%% Fig:D1 Integrability%%%%%%%%%%%%%%%%%%%%%%%%%%%%%
\begin{figure}[H]
	\centering
	\includegraphics[height=3.5cm,width=8cm]{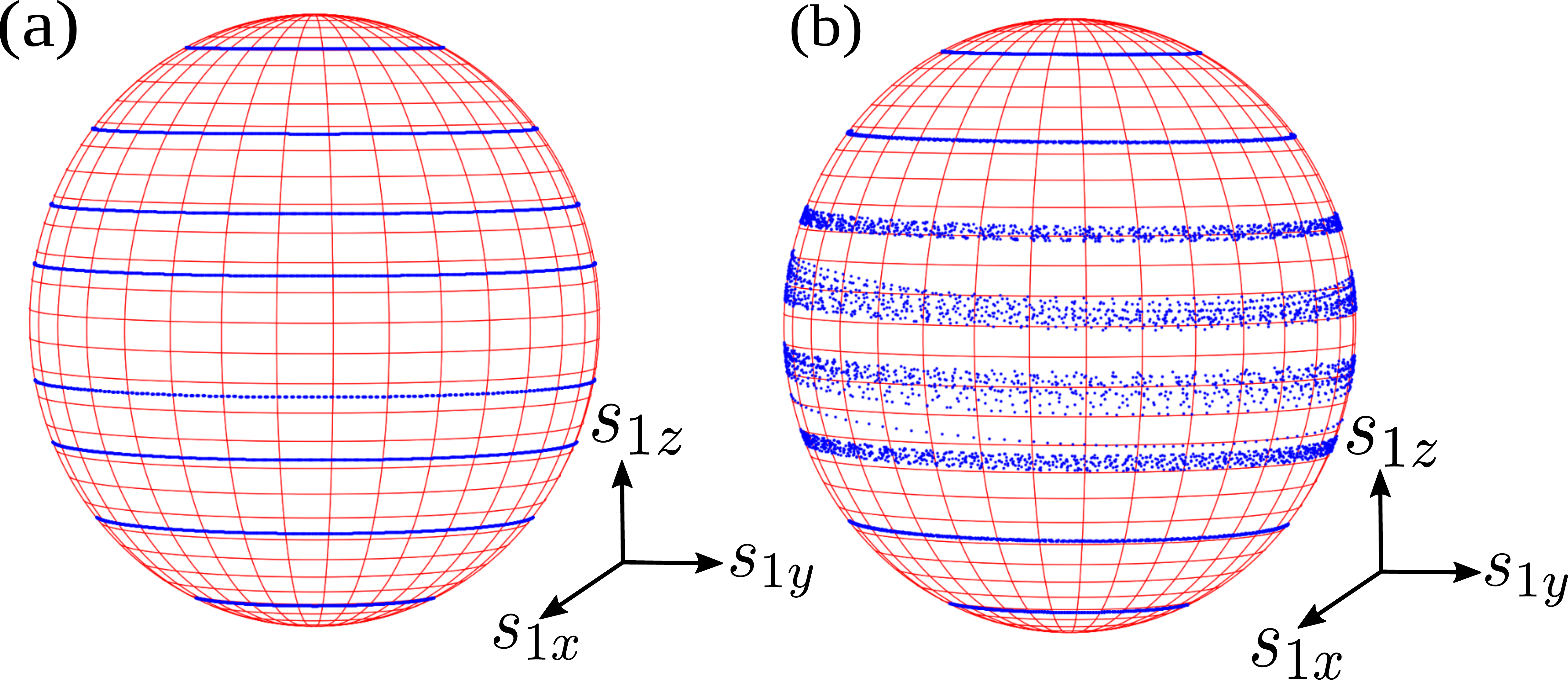}
	\caption{Comparison of phase space trajectories on Bloch sphere corresponding to a particular spin sector for (a) $\epsilon=0$ and (b) $\epsilon=0.01$, exhibiting diffusive behavior of the trajectories near the equator with $E\approx 0$, as a result of deviation from ergodicity for $\epsilon > 0$. 
}
	\label{Integrable limit}
\end{figure}
%%%%%%%%%%%%%%%%%%%%%%%%%%%%%%%%%%%%%%%%%%%%%%%%%%%%%%%%%%%%%%
For $\epsilon =0$, the classical Hamiltonian $\mathcal{H}_{cl}$ given in Eq.\eqref{semiclassical_ham_integrable} is  independent of the angle variables $\phi_i$, and the EOM yields the solution
$\phi_i = z_it + c$ with constant $z_i$, which represents precession of spins around the $z$-axis as depicted in Fig.\ref{Integrable limit}(a). As $\epsilon$ increases, the dynamics near the equator of the Bloch sphere corresponding to the energy $E\approx 0$ becomes diffusive, whereas regular trajectories are observed near the poles (see Fig.\ref{Integrable limit}(b)). 
The above results obtained from classical analysis at large $\mu$ are also supported by quantum analysis as well. To see this, we sort the energy levels obtained from exact diagonalization of the above Hamiltonian (Eq.(\ref{integrable_limit})) in ascending order, belonging to a particular parity as well as exchange symmetry sector (mentioned in main text). Next, we compute the level spacing $\delta_{\nu}=\mathcal{E}_{\nu}-\mathcal{E}_{\nu-1}$, which is shown with increasing value of the index $\nu$ in Fig.\ref{label spacing large mu}(a).
For the integrable limit with $\epsilon=0$, the level spacings are distributed in a regular fashion compared to the that for $\epsilon\neq0$. However, for $\epsilon \ll 1$, the level spacings corresponding to very low as well as high energy states show almost regular structure, whereas those at the middle of the spectrum ($E\approx 0$) exhibits significant deviation, which is consistent with the classical picture (see Fig.\ref{label spacing large mu}(a)). Also, such deviation increases as the Hamiltonian differs from the integrable limit with increasing value of $\epsilon$. A similar analysis has also been done in the context of Dicke model \cite{Emary_Brandes}.
%%%%%%%%%%%%%%%%%%%%%%%% Fig: D2 LEVEL SPACING %%%%%%%%%%%%%%%%%%%%%%%
%%%%%%%%%%%%%%%%%%%%%%%%%%%%%%%%%%%%%%%%%%%%%%%%%%%%%%%%%%%%%%
To confirm the crossover from chaotic to regular dynamics for large value of $\mu$ ($\epsilon\ll 1$), we compute the average ratio of consecutive level spacing $\langle r_{\nu}\rangle$ as a function of decreasing $\epsilon$, which exhibits an approach to Poissonian distribution, as depicted in Fig.\ref{label spacing large mu}(b).

\end{document}